\documentclass[showpacs,preprintnumbers,amsmath,amssymb,APSl,prd,nofootinbib,superscriptaddress, twocolumn]{revtex4-1}

\usepackage{dcolumn}
\usepackage{bm}
\usepackage{ifpdf}
\usepackage{hyperref}
\usepackage{bm}
\usepackage{xcolor,color,graphicx,graphics}
\usepackage[spanish,english]{babel}
\usepackage[latin1]{inputenc}
\usepackage[OT1]{fontenc}
\usepackage{latexsym,amssymb,amsmath,amsfonts}
\usepackage{makeidx}
\usepackage{graphicx}
\usepackage{epsfig}
\usepackage{natbib}
\usepackage{epstopdf}
\usepackage{mathrsfs}
\usepackage{hyperref}
\hypersetup{colorlinks=true, linkcolor=blue, citecolor=green}
\usepackage{enumerate}

\begin{document}
\title{Boson stars in Palatini $f(\mathcal{R})$ gravity}
\author{Andreu Mas\'o-Ferrando} \email{andreu.maso@uv.es}
\affiliation{Departamento de F\'{i}sica Te\'{o}rica and IFIC, Centro Mixto Universidad de Valencia - CSIC.
Universidad de Valencia, Burjassot-46100, Valencia, Spain}
\author{Nicolas Sanchis-Gual} \email{nicolas.sanchis@tecnico.ulisboa.pt}
\affiliation{Departamento  de  Matem\'{a}tica  da  Universidade  de  Aveiro  and  Centre  for  Research  and  Development in  Mathematics  and  Applications  (CIDMA),  Campus  de  Santiago,  3810-183  Aveiro,  Portugal} 

\author{Jos\'e A. Font} \email{j.antonio.font@uv.es}
\affiliation{Departamento de Astronom\'{\i}a y Astrof\'{\i}sica, Universitat de Val\`encia,
Dr. Moliner 50, 46100, Burjassot (Val\`encia), Spain}
\affiliation{Observatori Astron\`omic, Universitat de Val\`encia, Catedr\'atico 
 Jos\'e Beltr\'an 2, 46980, Paterna (Val\`encia), Spain}
	\author{Gonzalo J. Olmo} \email{gonzalo.olmo@uv.es}
	\affiliation{Departamento de F\'{i}sica Te\'{o}rica and IFIC, Centro Mixto Universidad de Valencia - CSIC.
Universidad de Valencia, Burjassot-46100, Valencia, Spain}
\affiliation{Departamento de F\'isica, Universidade Federal da
Para\'\i ba, 58051-900 Jo\~ao Pessoa, Para\'\i ba, Brazil}

\date{\today}
\begin{abstract}
  
We explore equilibrium solutions of spherically symmetric boson stars in the Palatini formulation of $f(\mathcal{R})$ gravity. We account for the modifications introduced in  the  gravitational sector  by using a recently established correspondence between modified gravity with scalar matter and general relativity with modified scalar matter. 
We focus on the quadratic theory $f(\mathcal{R})=R+\xi R^2$ and compare its solutions with those found in general relativity, exploring both positive and negative values of  the coupling parameter $\xi$. As matter source, a complex, massive scalar field with and without self-interaction terms is considered. Our results show that the existence curves of boson stars in Palatini $f(\mathcal{R})$ gravity are fairly similar to those found in general relativity. Major differences are observed for negative values of the coupling parameter which results in a repulsive gravitational component for high enough scalar field density distributions. Adding self-interactions makes the degeneracy between $f(\mathcal{R})$ and general relativity even more pronounced, leaving very little room for observational discrimination between the two theories. 
\end{abstract}

\maketitle

\section{Introduction}

The Advanced LIGO and Advanced Virgo observations of gravitational waves produced by the mergers of massive compact objects is providing crucial new information about the abundance and mass spectrum of such astrophysical entities~\cite{GWTC-1,GWTC-2}. 
New black hole populations, with  significantly larger masses than those of the black holes inferred from electromagnetic observations, have been unveiled, which challenges our theories of binary-black-hole (BBH) formation and raises questions about their origin. Most interestingly, the mass range of the BBH systems reported in~\cite{GWTC-2} has, on the one hand, begun to populate the gap between the heaviest neutron stars known and the black hole population (with the event  GW190814~\cite{GW190814}) and, on the other hand, in the high-mass end, it includes a system (GW190521) with components inside the pair-instability supernova gap, challenging our current models of stellar evolution~\cite{GW190521}.
Making sense of the data 
is creating tension between theoretical expectations and what data indicate. 
In particular, objects with 50-100 $M_\odot$ leading to mergers within the domain of intermediate black hole masses~\cite{GW190521} pose fundamental questions about the nature and origin of the merging compact objects (see~\cite{GW190521b,Deluca,Juan}). Although black holes of the Kerr type lead the bets as the primary sources of those mergers, it is an open issue how they could form and grow to the observed sizes. Moreover, current observations still cannot rule out the possibility that mergers of other types of dark compact objects, commonly referred to as exotic compact objects (ECOs) (see \cite{Cardoso-Pani} and references therein) could have led to the observed signals~\cite{Deluca,Juan}.  
Further evidence of such degeneracy  has also recently been reported from the {\it electromagnetic} channel~\cite{imitation}.

 Among  ECOs boson stars occupy a prominent role. These self-gravitating objects are  gravitationally  bound  configurations of scalar bosonic particles minimally coupled to Einstein's gravity, and their masses and sizes  range from atomic to astrophysical scales, depending on the mass of the bosonic particle. One should bear in mind that the scalar fields engendering such solutions need not represent stable fundamental bosonic fields gravitationally bound. In fact, they could be the result of complex effective interactions of different (more) fundamental fields in equilibrium, such as in Bose-Einstein condensates, thus allowing for the existence of objects much denser than typical neutron stars.  Since the seminal works of Kaup~\cite{Kaup} and Ruffini and Bonnazola~\cite{RB} boson stars have received significant attention and their stability properties and dynamics have been  investigated to a great extent (see \cite{Palenzuela:2017xcv,Fabrizio1} and references therein). It is thus well established that realistic  configurations of (rotating and non-rotating) compact objects of many solar masses can be generated numerically with appropriate model parameters, which allows to accommodate currently available observations without strictly requiring the existence of an event horizon. In fact, within the simplification of spherical symmetry boson stars in the ground state (i.e.~the fundamental family) have been shown to form dynamically from a dilute cloud of scalar particles through the  gravitational cooling mechanism~\cite{SeidelSuen} and to be stable under perturbations~\cite{Gleiser,Lee,Nico2017,Guzman}. 
Spherical boson stars have also been used  to  build  orbiting  binaries, which has allowed to compare their  gravitational-wave emission to that of black hole binaries~\cite{Palenzuela2007,Palenzuela2008}. Beyond spherical symmetry, equilibrium sequences of axisymmetric,  rotating boson stars have  been constructed in~\cite{Schunck,Yoshida,Kleihaus} and  simulations of binary mergers of rotating boson stars have been discussed by~\cite{Bezares2017,Palenzuela2017}.

A few recent studies have considered {\it vector} boson stars (i.e.~Proca stars~\cite{Brito})  finding  interesting similarities and some striking differences with their scalar counterparts. As in the scalar case spinning (and also non-rotating)  Proca stars  can  form  from  the  gravitational  collapse  of  a  dilute  cloud  of Proca  particles  with  non-zero  angular  momentum,  via  gravitational  cooling~\cite{Nicoletter, Fabrizio1} (see also~\cite{Fabrizio2} for the spherically-symmetric case).   
Rotating scalar boson stars were shown in~\cite{Nicoletter,Fabrizio1} to be transient objects, as they develop a non-axisymmetric instability which triggers the loss of angular momentum and their migration to the $m=0$, non-rotating family. Such behaviour is not found for fundamental ($m=1$) rotating Proca stars but it is nevertheless present for $m=2$ Proca stars which are observed to  migrate to the stable  spheroidal family. We note that recently~\cite{Siemonsen} have found situations when adding nonlinear interactions to the scalar potential quenches the non-axisymmetric instability of rotating boson stars discussed in~\cite{Nicoletter,Fabrizio1}. It is also worth mentioning that the direct detection of bosonic fields as constituents of Proca stars has been recently  proposed  in  connection  with  GW190521~\cite{Juan}.

The hypothetical existence of ECOs able to reach larger ranges of densities may open new opportunities to explore potential modifications of the gravitational sector in the strong-field regime. In fact, the absence of a horizon could make the innermost regions of those objects accessible to observation, potentially offering new insights on how to extend Einstein's gravity in the ultraviolet. In this sense, it is of interest to explore how structural properties such as mass and radius of boson stars could be affected by a modification of the gravitational Lagrangian  \cite{Olmo:2019flu,Rubiera-Garcia:2020gcl}. Given that $f(\mathcal{R})$ theories  \cite{DeFelice:2010aj,Olmo:2011uz,Harko:2018ayt} offer a large amount of freedom while keeping the field equations within reasonable limits of simplicity, in this work we will explore the impact that high-energy modifications of the gravitational interaction of the $f(\mathcal{R})$ type could have on the astrophysical properties of boson stars. {Similar studies have already been carried out in other theories of gravity, such as in scalar-tensor theories \cite{Torres:1997cvs}, Horndeski theories \cite{Brihaye:2016lin,Verbin:2017bdo},  and theories with Gauss-Bonnet couplings \cite{Hartmann:2013tca,Brihaye:2013zha,Baibhav:2016fot}, among others.}

When the fundamental nature of the gravitational sector is relaxed and $f(\mathcal{R})$ extensions are allowed, one must face the question of how to derive the field equations. The traditional approach assumes that the space-time is a Riemannian structure completely described by the metric tensor. But one can also consider a metric-affine geometry with {\it a priori} independent metric and affine structures \cite{Hehl:1994ue,Olmo:2011uz,BeltranJimenez:2019tjy} {(for an analysis of boson stars in theories with torsion of the $f(T)$ type, see \cite{Ilijic:2020vzu})}. 
Given that we have no compelling evidence about the kind of geometry associated to the space-time, it seems fair not to discriminate any of (at least) those options {\it a priori}. Though, in practice,  this choice has no impact in the case of GR minimally coupled to scalar fields, when $f(\mathcal{R})$ or more general extensions are considered the difference is certainly relevant, leading to two inequivalent sets of equations\footnote{In fact, the equivalence only seems to hold for the so-called Lovelock theories \cite{Borunda:2008kf}. }. In the usual metric approach, the nonlinearity of the $f(\mathcal{R})$ Lagrangian induces the emergence of a dynamical scalar degree of freedom in the gravitational sector \cite{Olmo:2006eh}. The resulting theory turns out to be equivalent to a particular case of Brans-Dicke scalar-tensor theory, with the Brans-Dicke parameter $\omega=0$. In the metric-affine (or Palatini) case, a scalar-tensor representation is also possible, with $\omega=-3/2$, though in this case the new scalar degree of freedom is not dynamical  \cite{Olmo:2005zr,Olmo:2005hc}. The effect of the Palatini $f(\mathcal{R})$ Lagrangian is to induce non-linearities in the matter sector. As a result, the vacuum field equations of Palatini $f(\mathcal{R})$ theories exactly recover those of GR with an effective cosmological constant regardless of the $f(\mathcal{R})$ function chosen. This sharply contrasts with the metric formulation, in which the space-time is generically curved even in the absence of sources, which can lead to long-range effects in the Newtonian/post-Newtonian regimes (depending on model parameters) and induces an extra polarization mode in the spectrum of gravitational waves. The Palatini formulation, instead, predicts only two polarizations which propagate at the speed of light in vacuum, making them consistent with current constraints coming from neutron stars mergers \cite{Lombriser:2015sxa,Lombriser:2016yzn,Baker:2017hug,Sakstein:2017xjx,Creminelli:2017sry,Ezquiaga:2017ekz,Ezquiaga:2018btd}. 

The Palatini formulation of $f(\mathcal{R})$ theories and of other theories based on the Ricci tensor turns out to be particularly interesting from a computational point of view. In the Palatini framework it is possible to transform the  problem of a modified gravity theory minimally coupled to a scalar field (or other matter source) into a standard problem in GR minimally coupled to a modified scalar Lagrangian (or other matter Lagrangian) \cite{Afonso:2018hyj,Delhom:2019zrb,Afonso:2018mxn,Afonso:2018bpv}. This property has been used recently to generate new analytical solutions for static, spherically symmetric scalar compact objects in Palatini $f(\mathcal{R})$ and other theories \cite{Afonso:2019fzv}. In fact, starting from a known solution of GR with a spherically symmetric, static, massless real scalar field, which represents a naked singularity, it was possible to obtain new exotic compact objects such as wormholes and other configurations with peculiar causal properties within the high density region but almost identical to the standard GR solution in its exterior regions (where the energy density rapidly drops to zero and the dynamics tends to that of GR). New exact rotating solutions \cite{Guerrero:2020azx,Shao:2020weq} and even multicenter solutions \cite{Olmo:2020fnk} (without defined symmetry) have been constructed using this approach. 

Here we will take advantage of this correspondence between modified gravity with scalar matter and GR with modified scalar matter  to implement numerically the analysis of boson stars governed by a specific Palatini $f(\mathcal{R})$ Lagrangian. Regular solutions of this type require the use of a complex, massive scalar field with a harmonic time dependence. The introduction of a mass  breaks the shift symmetry of the massless case in such a way that no exact analytical solutions are known. In addition, the harmonic time dependence adds new difficulties in the system of equations that can only be solved by resorting to numerical methods. For concreteness, in this work we will focus on the quadratic theory $f(\mathcal{R})=R+\xi R^2$. In order to use standard numerical methods, we will use the correspondence mentioned above to determine the modified scalar field Lagrangian that coupled to GR can be used to generate the solutions of the original $f(\mathcal{R})$ plus scalar theory. As we shall see, the solutions obtained indicate that spherically symmetric, stationary boson stars are robust under modifications of the gravitational dynamics for a broad range of the gravitational coupling $\xi$. Our results also provide information about the conditions needed to generate wormhole solutions. 

This work is organized as follows: Section~\ref{framework} deals with the mathematical framework of our study. It   discusses the correspondence between modified gravity and GR and presents the system of differential equations to solve  to construct boson star models in spherical symmetry in the absence of self-interactions. The corresponding numerical framework is discussed in Section~\ref{numerical}. Our results are presented in Section~\ref{results}. Here, we also discuss the modifications in the equations to allow for a self-interaction term in the Klein-Gordon potential and draw comparisons between our numerical solutions with and without self-interactions.
Finally, Section~\ref{conclusions} summarizes our main findings. Unless stated otherwise we use a system of natural units in which $c=G=\hbar=1$. 

\section{Correspondence with GR and field equations}
\label{framework}

We will be dealing with a theory of the form 
\begin{equation}\label{eq:Ac1}
S_{f(\mathcal{R})}=\int d^4x \sqrt{-g} \frac{f(\mathcal{R})}{2 \kappa} -\frac{1}{2}\int d^4x \sqrt{-g} P(X,\Phi) \quad .
\end{equation}
where gravity is described in terms of a Palatini $f(\mathcal{R})$ function and the matter sector is represented by a complex scalar field $\Phi$ with Lagrangian $P(X,\Phi)=X-2V(\Phi)$, where $X=g^{\alpha \beta}\partial_{\alpha}\bar{\Phi}\partial_{\beta}\Phi$, $V(\Phi)=-\mu^2 \bar{\Phi} \Phi/2$, and $\mu$ is the scalar field mass. Here we are defining $\mathcal{R}=g^{\mu\nu}R_{\mu\nu}(\Gamma)$, with $R_{\mu\nu}(\Gamma)$ representing the Ricci tensor of a connection $\Gamma^\lambda_{\alpha\beta}$ a priori independent of the metric $g_{\mu\nu}$. Manipulating the field equations that follow from independent variations of the metric and the connection, one finds that the explicit relation between  $\Gamma^\lambda_{\alpha\beta}$ and $g_{\mu\nu}$ is given by 
\begin{equation}
\Gamma^\lambda_{\mu\nu}=\frac{q^{\lambda\rho}}{2}\left[\partial_\mu q_{\rho\nu}+\partial_\nu q_{\rho\mu}-\partial_\rho q_{\mu\nu}\right] \ ,
\end{equation}

where we have introduced 
\begin{equation}\label{eq:conformal}
q_{\mu\nu}\equiv f_{\mathcal{R}} g_{\mu\nu} \ ,
\end{equation}
with $f_\mathcal{R}\equiv \partial f/\partial \mathcal{R}$. We note that the conformal factor $f_{\mathcal{R}}$ must be regarded as a function of the metric $g_{\alpha\beta}$ and the matter fields which is specified by the algebraic equation
\begin{equation}\label{eq:Trace}
\mathcal{R}f_{\mathcal{R}}-2f=\kappa T \ ,
\end{equation}
where $T$ represents the trace of the matter stress-energy tensor, which is defined as
\begin{equation}\label{eq:Tmn-g}
{T}_{\mu \nu}\equiv-\frac{2}{\sqrt{-q}}\frac{\delta(\sqrt{-q} P(X,\Phi))}{\delta g^{\mu \nu}} \ .
\end{equation}
 For simplicity, we will specify the gravity Lagrangian by the quadratic function
\begin{equation}
f(\mathcal{R})=\mathcal{R}+\xi \mathcal{R}^2 \ ,
\end{equation}
which inserted in (\ref{eq:Trace}) leads to the relation $\mathcal{R}=-\kappa T$, exactly like in GR.  
We will refer to the representation (\ref{eq:Ac1}) of the theory as the $f(\mathcal{R})$ frame. Note that in this frame the scalar $\Phi$ is minimally coupled to the metric $g_{\mu\nu}$.

 As it was shown in \cite{Afonso:2018hyj}, there exists a correspondence between the theory (\ref{eq:Ac1}) and the Einstein-Hilbert action of the metric $q_{\mu\nu}$ minimally coupled to a matter Lagrangian $K(Z,\Phi)$ (from now on the Einstein frame), namely,  
\begin{equation}
S_{EH}=\int d^4x \sqrt{-q} \frac{R}{2 \kappa} -\frac{1}{2}\int d^4x \sqrt{-q} K(Z,\Phi) \quad ,
\end{equation}
where the kinetic term $Z=q^{\alpha \beta}\partial_{\alpha}\bar{\Phi}\partial_{\beta}\Phi$ is now contracted with the (inverse) metric $q^{\alpha \beta}$ and $R$ is the Ricci scalar of the metric  $q_{\alpha \beta}$, i.e., $R= q^{\alpha \beta}R_{\alpha \beta}(q)$.

For the specified $f(\mathcal{R})$ and $P(X,\Phi)$ functions it can be shown that \cite{Afonso:2018hyj}
\begin{equation}\label{eq:K}
K(Z,\Phi)=\frac{Z-\xi \kappa Z^2}{1-8 \xi \kappa V}-\frac{2V}{1-8 \xi \kappa V} \quad .
\end{equation}
As we can see, non-linearities in the gravitational sector of the $f(\mathcal{R})$ frame have been transferred into non-linearities in the matter sector of the Einstein frame. Because of this relation between frames, in order to solve the field equations of $f(\mathcal{R})$ gravity coupled to a scalar field  we will solve instead the corresponding problem in GR coupled to the non-linear scalar field matter Lagrangian (\ref{eq:K}). Once the metric $q_{\mu\nu}$ and the scalar field $\Phi$ have been found, we automatically have the metric $g_{\mu\nu}$ via the conformal relation (\ref{eq:conformal}). \\

To proceed, we will now consider the Einstein-Klein-Gordon system in the Einstein frame. The corresponding stress-energy tensor is given by
\begin{eqnarray}
\tilde{T}_{\mu \nu}&\equiv&-\frac{2}{\sqrt{-q}}\frac{\delta(\sqrt{-q} K(Z,\Phi))}{\delta q^{\mu \nu}}\nonumber \\
&=&\frac{1}{2(1+4 \xi \kappa \mu^2 |\Phi|^2)}\left[\left(\partial_{\mu}\bar{\Phi}\partial_{\nu}\Phi +\partial_{\nu}\bar{\Phi}\partial_{\mu}\Phi\right)\left(1-2 \xi \kappa Z\right) \right.\nonumber \\ 
&-&\left.q_{\mu \nu}\left(\partial^{\alpha}\bar{\Phi}\partial_{\alpha}\Phi\left(1- \xi \kappa Z \right) +\mu^2 |\Phi|^2 \right)\right]  \ ,
\end{eqnarray}
which should not be confused with the $T_{\mu\nu}$ defined in (\ref{eq:Tmn-g}). Following the approach of \cite{Herdeiro:2017xcv,Palenzuela:2017xcv}, we will consider spherical  stars described by a scalar-field profile of the form $\Phi(x,t)=\phi(x)e^{i \omega t}$, where $\omega$ is the oscillation frequency of the field. Since we are describing spherically symmetric configurations we will use polar-areal coordinates. Our ansatz for the metric $g_{\mu\nu}$ is 
\begin{equation}\label{eq:ds2fR}
ds_{f}^2=-A^2(r) dt^2+B^2(r)dr^2+r^2d\theta^2+r^2 \sin^2 {\theta}d\varphi^2\; .
\end{equation}
Analogously, for the metric $q_{\mu\nu}$ we take
\begin{equation}\label{eq:ds2GR}
ds_{\rm GR}^2=-\alpha^2(x) dt^2+\beta^2(x)dx^2+x^2d\theta^2+x^2 \sin^2 {\theta}d\varphi^2\; .
\end{equation}
From the Einstein equations associated with the line element (\ref{eq:ds2GR}), the components $G_{tt}$ and $G_{xx}$ lead to
\begin{eqnarray}\label{beta}
\frac{\partial_{x}\beta}{\beta}&=&\frac{1-\beta^2}{2x}+\frac{1}{1+4 \xi \kappa \mu^2 \phi^2 }\frac{\kappa x}{4} \left\lbrace\frac{}{} \mu^2 \beta^2  \phi^2 \right.\nonumber \\
&+&\left.\left( \omega^2\phi^2\frac{ \beta^2  }{ \alpha^2}+ \psi^2 \right)\left(1-2 \kappa \xi\left( -\frac{\omega^2   \phi^2}{ \alpha^2} +\frac{\psi^2}{ \beta^2} \right)\right) \right.\nonumber \\
& +&\left.2\kappa \xi \beta^2\left( \frac{\omega^2   \phi^2}{ \alpha^2} -\frac{\psi^2}{ \beta^2} \right)^2  \right\rbrace ,   \\
\frac{\partial_{x}\alpha}{\alpha}&=&\frac{\beta^2-1}{x}+\frac{\partial_{x}\beta}{\beta}+\frac{1}{1+4 \xi \kappa \mu^2 \phi^2 }\frac{\kappa x}{4}\left\lbrace
-2 \mu^2 \beta^2  \phi^2 \right.\nonumber \\
&-&\left.2\kappa \xi \beta^2\left( \frac{\omega^2   \phi^2}{ \alpha^2} -\frac{\psi^2}{ \beta^2} \right)^2\right\rbrace , \label{alpha}
\end{eqnarray}
where the quantity 
\begin{equation}\label{phi}
\psi\equiv \partial_{x}\Phi \,,
\end{equation}
satisfies
\begin{eqnarray}\label{psi}
\partial_{x}\psi&=&\frac{1}{(1+4 \xi \kappa \mu^2 \phi^2 )\left[1-2 \xi \kappa\left( -\frac{\omega^2 \phi^2 }{\alpha^2}+  \frac{3\psi^2}{\beta^2}\right)\right]} \left\lbrace\frac{}{} \right.\nonumber\\ 
&-&\psi\left(\frac{2}{x}+\frac{\partial_{x}\alpha}{\alpha}- \frac{\partial_{x}\beta}{\beta} \right)\left(1+4\kappa \xi \mu^2\phi^2\right)
\nonumber\\
&\times&\left[1-2 \xi \kappa\left( -\frac{\omega^2 \phi^2 }{\alpha^2}+  \frac{\psi^2}{\beta^2}\right)\right] \nonumber\\
 &-&\omega^2 \phi \frac{\beta^2}{\alpha^2}\left[1+2 \xi \kappa\left( \frac{\omega^2 \phi^2 }{\alpha^2}+  \frac{\psi^2}{\beta^2}\right)\right]\nonumber\\
&+&\beta^2\phi \mu^2 \left(1+4\kappa \xi \frac{\psi^2}{\beta^2}\right)\nonumber\\
&+&\kappa \xi \left[ \frac{4 \omega^2 \phi^2 \psi}{\alpha^2}\frac{\partial_{x}\alpha}{\alpha}\left(1+4\kappa\xi\mu^2\phi^2\right)\right.\nonumber\\
&-&\left.\frac{4\psi^3}{\beta^2}\frac{\partial_{x}\beta}{\beta}\left(1+4\kappa \xi \mu^2\phi^2\right)\right]\nonumber\\
&-&\left.4\kappa^2 \xi^2 \mu^2 \phi \beta^2\left(\frac{\omega^4\phi^4}{\alpha^4}+\frac{3 \psi^4}{\beta^4}\right) \right\rbrace \ .
\end{eqnarray}
The above four equations \eqref{beta}-\eqref{psi} form the EKG system that we need to solve to build spherically symmetric boson star models in a quadratic $f(\mathcal{R})$ theory. For future reference, it is convenient to write explicitly the form of the conformal factor $f_{\mathcal{R}}$ as follows:
\begin{eqnarray}
f_{\mathcal{R}}&=&1+2\xi R=1+2\xi \kappa(X-4V) \nonumber \\
&=&1+2\xi \kappa \left[-\frac{\omega^2 \phi^2}{A^2}+\frac{\psi^2}{B^2}+2 \mu^2 \phi^2\right]\ .
\end{eqnarray} 
Note that using the relation $X=f_\mathcal{R} Z$ we can also write the conformal factor in terms of the $\text{GR}-$frame variables
\begin{eqnarray}\label{f_Rx}
f_\mathcal{R}&=&1+2\xi\kappa\left[\frac{\left(1-8\xi \kappa V\right)Z}{1-2\xi \kappa Z}-4V\right]\nonumber \\
&=&\frac{1+4\xi \kappa \mu^2 \phi^2}{1-2\xi \kappa \left(-\frac{\omega^2 \phi^2}{\alpha^2}+\frac{\psi^2}{\beta^2}\right)} \ .
\end{eqnarray}\\

\section{Numerical analysis}
\label{numerical}
\subsection{Boundary Conditions}\label{subsec_bc}
In order to solve this system of differential equations suitable boundary conditions have to be provided. We will regard the $f(\mathcal{R})$ frame as the frame in which the physical boundary conditions must be specified. Accordingly, we impose asymptotic flatness at infinity and regularity at the origin for the line element (\ref{eq:ds2fR}), which translates into 
\begin{eqnarray}
\phi(\infty)&=&0 \ , \  \psi(\infty)=0 \ , \  B^2(\infty)=1 \ , \  
A^2(\infty)=1 \\
\phi(0)&=&\phi_0  \ , \   \psi(0)=0 \ , \  
\partial_{r} B^2(0)=0 \ , \  
\partial_{r} A^2(0)=0 \ .\nonumber 
\end{eqnarray}
The asymptotic flatness condition leads us to $f_\mathcal{R}\rightarrow 1$ when $r\rightarrow \infty$. Recalling the conformal relation \eqref{eq:conformal}, we can re-express the above conditions in the Einstein frame variables. First, we take a look at the area of the 2-spheres, which are related according to $x^2=f_\mathcal{R} r^2$. If one assumes that $f_\mathcal{R} \neq 0$ everywhere then it follows that $x\to 0$ when $r\to0$, so the boundary conditions read as
\begin{eqnarray}
\phi(\infty)&\equiv&\phi(x(r))|_{r=\infty}=0 \,, \\
\psi(\infty)&\equiv&\psi(x(r))|_{r=\infty}=0 \,,    \\
\beta^2(\infty)&\equiv&\beta^2(x(r))|_{r=\infty}=f_\mathcal{R}(\infty) B^2(\infty)=1 \,,   \\ 
\alpha^2(\infty)&\equiv&\alpha^2(x(r))|_{r=\infty}=f_\mathcal{R}(\infty) A^2(\infty)=1 \,, \ 
\\
\nonumber \\
\phi(0)&\equiv&\phi(x(r))|_{r=0}=\phi_0  \,, \\
\psi(0)&\equiv&\psi(x(r))|_{r=0}=0 \,, \ \label{eq:psi_0}\\
\label{beta_regularity}
\left[\partial_{x}\beta^2\right](0)&\equiv&\left[\partial_{x}\beta^2\right](x(r))|_{r=0} \nonumber \\
&=&2 \sqrt{f_\mathcal{R}}B \partial_{r}(\sqrt{f_\mathcal{R}}B)\partial_{x}r|_{r=0}\nonumber \\
&=&\frac{B^2}{\sqrt{f_\mathcal{R}}}\partial_{r}f_\mathcal{R}|_{r=0} =0 \,, \\
\label{alpha_regularity}
\left[\partial_{x}\alpha^2\right](0)&\equiv&\left[\partial_{x}\alpha^2\right](x(r))|_{r=0}\nonumber\\
&=&2 \sqrt{f_\mathcal{R}}A \partial_{r}(\sqrt{f_\mathcal{R}}A)\partial_{x}r|_{r=0}\nonumber \\
&=&\frac{A^2}{\sqrt{f_\mathcal{R}}}\partial_{r}f_\mathcal{R}|_{r=0}=0 \ .
\end{eqnarray}
Substituting Eqs.~\eqref{beta_regularity} and \eqref{alpha_regularity} into Eqs.~\eqref{beta} and \eqref{alpha} respectively leads to $\beta^2(0)=1$, $\alpha^2(0)=\alpha_{0}^2$. This puts forward that the assumption of asymptotic flatness and regularity at the origin in the $f(\mathcal{R})$ frame implies the same conditions in the Einstein frame. \\

\subsection{Scaling and dimensionless quantities}
In order to absorb some parameters to deal with dimensionless expressions, let us perform a re-scaling of the system as $r \rightarrow \mu r $, $t\rightarrow \omega t$. The factor $\kappa$ from the Einstein field equations can be absorbed by a redefinition of the matter fields
\begin{equation}
\phi \rightarrow \sqrt{\frac{2}{\kappa^2}} \phi \quad , \quad \psi \rightarrow \sqrt{\frac{2}{\kappa^2}} \psi \quad ,
\end{equation}
which leaves the scaled matter fields dimensionless. Using the symmetry of the equations of motion we can set 
\begin{equation}
 \alpha \rightarrow \frac{\omega}{\mu}\alpha \quad .
\end{equation}
When performing the numerics, we will set the field mass to $\mu=1$. We now introduce an expression for the Misner-Sharp mass
\begin{eqnarray}
M_{\text{MS}}&=&\frac{r_{\text{max}}}{2}  \left(1-\frac{1}{B^2(r_{\text{max}})}\right)\nonumber\\
&=&\frac{x_{\text{max}}}{2\sqrt{f_\mathcal{R}(x_{\text{max}})}}\left(1-\frac{f_\mathcal{R}(x_{\text{max}})}{\beta^2(x_{\text{max}})}\right)\\
&\approx& \frac{x_{\text{max}}}{2}\left(1-\frac{1}{\beta^2(x_{\text{max}})}\right)\quad ,
\end{eqnarray}
which gives us a numerical value for the mass related to the physical one by 
\begin{equation}\label{eq:M_MS}
M_{\text{MS}}=\frac{\mu M_{\text{phys}}}{M_{\text{Pl}}},
\end{equation}
where $M_{\text{Pl}}$ is the Planck mass. Note that the Misner-Sharp mass is the same in both frames. We introduce also an expression for the Noether charge, which arises from the global $U(1)$ symmetry $\Phi \rightarrow e^{i \varphi} \Phi$, and can be identified as the particle number 
\begin{equation}
\begin{aligned}\label{eq:N_fR}
N&=\int_{\Sigma}dV\sqrt{-g}g^{t \nu} \frac{i}{2}\left(\bar{\Phi}\partial_{\nu}\Phi-\Phi\partial_{\nu}\bar{\Phi}\right)\\
&=4\pi\int_{0}^{\infty}dr r^2 \omega \frac{\phi^2 B}{A}\\
&=4\pi\int_{0}^{\infty} \frac{dxx^2}{f_\mathcal{R}^{3/2}} \omega \frac{\phi^2 \beta}{\alpha} \quad ,\\
\end{aligned}
\end{equation}
and its relation with the physical value
\begin{equation}
N_{\text{phys}}=\frac{2 N}{\kappa \mu^2}\quad .
\end{equation}
The notion of binding energy arises naturally from the above definitions as 
\begin{equation}
E_{b}=M_{\text{MS}}-N\mu\quad ,
\end{equation}
and its sign will determine the stability of the boson star.\\

\subsection{Numerical method}
In order to solve numerically the re-scaled analogous EKG system with the provided boundary conditions we use a fourth-order Runge-Kutta scheme with adaptive stepsize. The conditions at the origin are evaluated at $x=10^{-6}$ in order to avoid indeterminations. Then an equidistant grid with spatial resolution $\Delta x=0.0025$ is used
and a global tolerance of $1.5\times10^{-14}$. Furthermore, for a given central value of the scalar field $\phi_0$ we have to adjust which frequency $\omega$ (integrated in $\alpha$ after the scaling) matches the desired asymptotic behavior. This is done by using a shooting method that integrates from the origin towards the outer boundary. There exists a set of $\omega^{(n)}$ values that satisfies this condition, and as $n$ increases also does the number of radial nodes of $\phi$. Here, we will focus on the nodeless $n=0$ case, known as the ground state or fundamental family.

From the scalings and redefinitions of parameters that we did in the previous section, the gravitational coupling $\xi$, with dimensions of length square, is now being measured in units of the inverse length defined by $\mu^2=1/l_\mu^2$, such that $\xi \mu^2$ is dimensionless. On physical grounds one expects  $\xi \mu^2\sim l_\xi^2/l_\mu^2 \ll 1$ but since we are mainly interested in a qualitative study of theories with positive and negative $\xi$, the coupling magnitude has been chosen large enough to easily notice the relevant features of each case. Hence, we choose $\xi \mu^2=-0.1,-0.05,-0.02,-0.01,0.01,0.1$ to explore $f(\mathcal{R})$ theories, and we note that $\xi=0$ is equivalent to GR. Absolute bounds on $\xi$ in the Palatini approach can be derived from the analysis of the weak-field limit presented in \cite{Olmo:2005zr}, leading to $|\xi|\ll 2\times 10^{12} $ cm$^2$. Another bound can be set by considering scenarios in which electric and gravitational (Newtonian) forces become of the same order of magnitude \cite{Avelino:2012qe,BeltranJimenez:2017doy}, leading to $|\xi|< 6 \times 10^{9} $ cm$^2$. By contrast, bounds on $\xi$ in the metric formalism range from $|\xi|< 5 \times 10^{15} $ cm$^2$ using Gravity Probe B data to $|\xi|< 1.7 \times 10^{18} $ cm$^2$ by analysing the precession of binary pulsars, while the E\"{o}t-Wash experiment yields $|\xi|<  10^{-6} $ cm$^2$ \cite{Naf:2010zy}.

\section{Results}
\label{results}
\subsection{Non-self-interacting scalar field}

\begin{figure*}[t!]
	\includegraphics[width=0.9\textwidth]{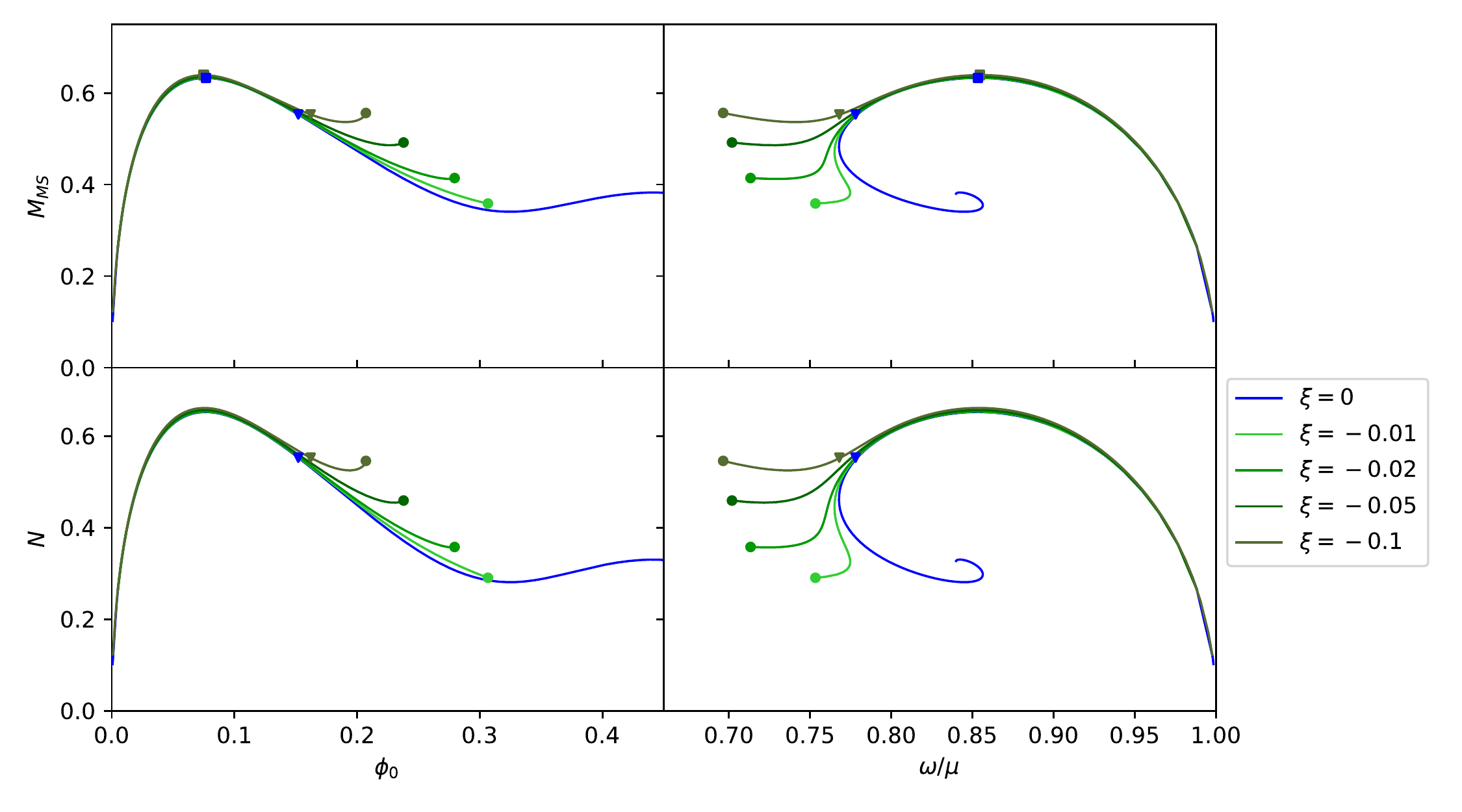}
	\includegraphics[width=0.9\textwidth]{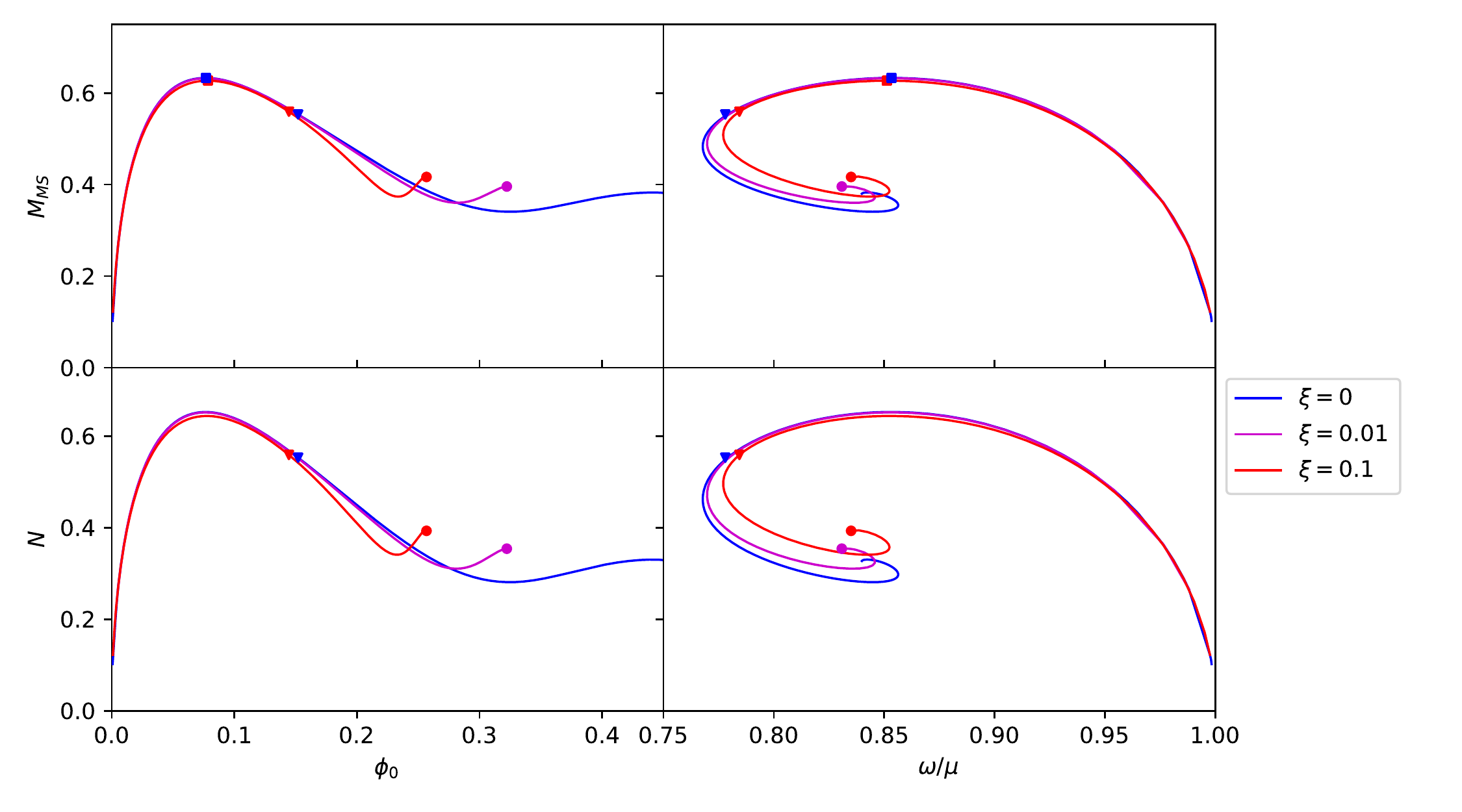}
	\caption{Existence curves of spherically symmetric boson stars in Palatini $f(\mathcal{R})$ gravity, obtained by solving  equations \eqref{beta}-\eqref{phi}. Solutions with $\xi<0$ ($\xi>0$) are plotted in the upper (lower) panel.  The left panels display the Misner-Sharp mass and the particle number $N$ against the central value of the scalar field $\phi_0$. The same quantities are plotted in the right panels against the frequency of the scalar field $\omega$ in units of $\mu$. Square symbols indicate the maximum values of the mass, inverted triangles the values where the binding energy changes sign, while circles signal the last solution we could build in $f(\mathcal{R})$. No circle is shown in the GR solutions as further solutions can be built. 
	}
	\label{fig:inidata}
\end{figure*}

\begin{table*}
	\centering
	\begin{tabular*}{0.8\textwidth}{|c @{\extracolsep{\fill}} | ccc|cc|c|ccc|}
		\hline
		$\xi$ & $M_{\text{max}}$	& $\phi_0(M_{\text{max}})$&$\omega(M_{\text{max}})$& $\phi_0(E_b=0)$&$\omega(E_b=0)$ & $\omega_{\text{min}}$ & $\phi_0^{\text{last}}$ & $\alpha_0^{\text{last}}(\mu/\omega)$ &$\omega^{\text{last}}$\\ \hline
		-0.10 & 0.6393 &0.0749&0.855& 0.1621&0.768&0.697&0.2071&0.4662&0.697\\
		-0.05& 0.6361 &0.0756&0.854& 0.1566&0.775&0.702&0.2380&0.3876&0.702\\
		-0.02& 0.6342 &0.0758&0.854& 0.1540&0.778&0.713&0.2795&0.2838&0.713\\
		-0.01& 0.6336 &0.0761&0.854& 0.1531&0.778&0.772&0.3066&0.2227&0.753\\
		0.0 & 0.6330 &0.0769&0.853& 0.1522&0.778&0.768&-&-&-\\
		0.01& 0.6323 &0.0784&0.853& 0.1515&0.780&0.769&0.3251&0.0844&0.831\\
		0.10 & 0.6270 &0.0784&0.851& 0.1445&0.784&0.777&0.2567&0.0357&0.835 \\
		\hline
	\end{tabular*}
	\caption{Values of the main physical quantities of our boson star models in Palatini $f(\mathcal{R})$ gravity. From left to right the columns report: 1st column: value of the gravitational coupling parameter; 2nd to 4th columns: parameters for maximal-mass solutions (squares in Fig.~\ref{fig:inidata}); 5th and 6th column: parameters of solutions with null binding energy (inverted triangles in Fig.~\ref{fig:inidata}); 7th column: value of minimal frequency in the mass-frequency plot, 8th to 10th column: parameter values for the last solution we are able to compute (circles in Fig.~\ref{fig:inidata}).}
	\label{tab:values_inidata}
\end{table*}

Figure \ref{fig:inidata} shows existence plots of various boson star solutions in the theories considered, both for positive and negative values of the coupling constant $\xi$.
Each solution is characterized by a given central scalar field amplitude $\phi_0$ and a frequency $\omega$, for which a mass $M_{\rm MS}$ and a particle number $N$ is computed. These solutions are consistent with the imposed boundary conditions, being regular at the origin and asymptotically flat. 

As one can see from Figure \ref{fig:inidata}, for small values of $\phi_0$ the solutions of  $f(\mathcal{R})$ theories are almost coincident with those of GR ($\xi=0$). However, for higher values of $\phi_0$ clear differences can be seen. This is due to the fact that the quadratic corrections in the gravitational sector become more relevant for high energy concentrations. The most notorious difference between the boson stars of GR and those of $f(\mathcal{R})$ is in their domain of existence. For GR, the interval shown does not exhibit any upper limit on the $\phi_0$ axis while this is not the case in $f(\mathcal{R})$.  
Similarly as in GR, boson stars in $f(\mathcal{R})$ theories exist in a bounded range of frequencies and masses, always with $\omega < \mu$. The case $\omega=\mu$ represents the limit where $M_{\rm MS}=0$. We also observe that as the gravitational coupling grows the maximum mass slightly decreases and the point signalling a vanishing value of the binding energy moves to lower values of the central scalar field amplitude and to larger values  of the oscillation frequency. 

Let us now focus on the $\xi<0$ case (top panels of Fig.~\ref{fig:inidata}). As the coupling parameter becomes more negative, the solutions depart more clearly from those of GR. The effect of the negative coupling is to generate a repulsive gravitational component when the scalar field density is high enough. This explains why slightly higher values of $M_{\text{max}}$ are allowed in these configurations, because a larger number of particles can be sustained due to the repulsive force. 

It is remarkable that at some point below $\xi<-0.01$, the dependence of  $M_{\text{MS}}$ with frequency departs from the well-known spiral behavior of GR and becomes a one-valued function. {Deviations from the spiral pattern are also observed in other theories of gravity as well \cite{Baibhav:2016fot,Brihaye:2016lin,Ilijic:2020vzu}, though in those cases solutions can be found over a larger range of frequencies.} This behavior is not observed for positive values of $\xi$, which still produce the same spiral pattern as GR.
In addition, between $\xi=-0.02$ and $\xi=-0.05$ the solutions start showing a local minimum for $M_{\text{MS}}$, which is not observed for positive values of $\xi$. Table \ref{tab:values_inidata} provides the values of $\phi_0$ and $\omega$ for a sample of our solutions, including those with the largest $\phi_0$ achievable, $\phi_0^{\rm{last}}$, represented by a solid circle in the plots of Fig.~\ref{fig:inidata}.  We note that we cannot compute numerically solutions beyond $\phi_0^{\rm{last}}$, which tends to $\phi^2_0 \omega^2/\alpha^2_0 = -1/(2 \xi \kappa ) $. Given the boundary condition (\ref{eq:psi_0}), this value of $\phi_0$ would make the conformal factor diverge at the origin (see Eq.~(\ref{f_Rx})), which precludes finding solutions in this region of the  parameter space. 

\begin{figure}[h]
	\centering
	\includegraphics[width=\linewidth]{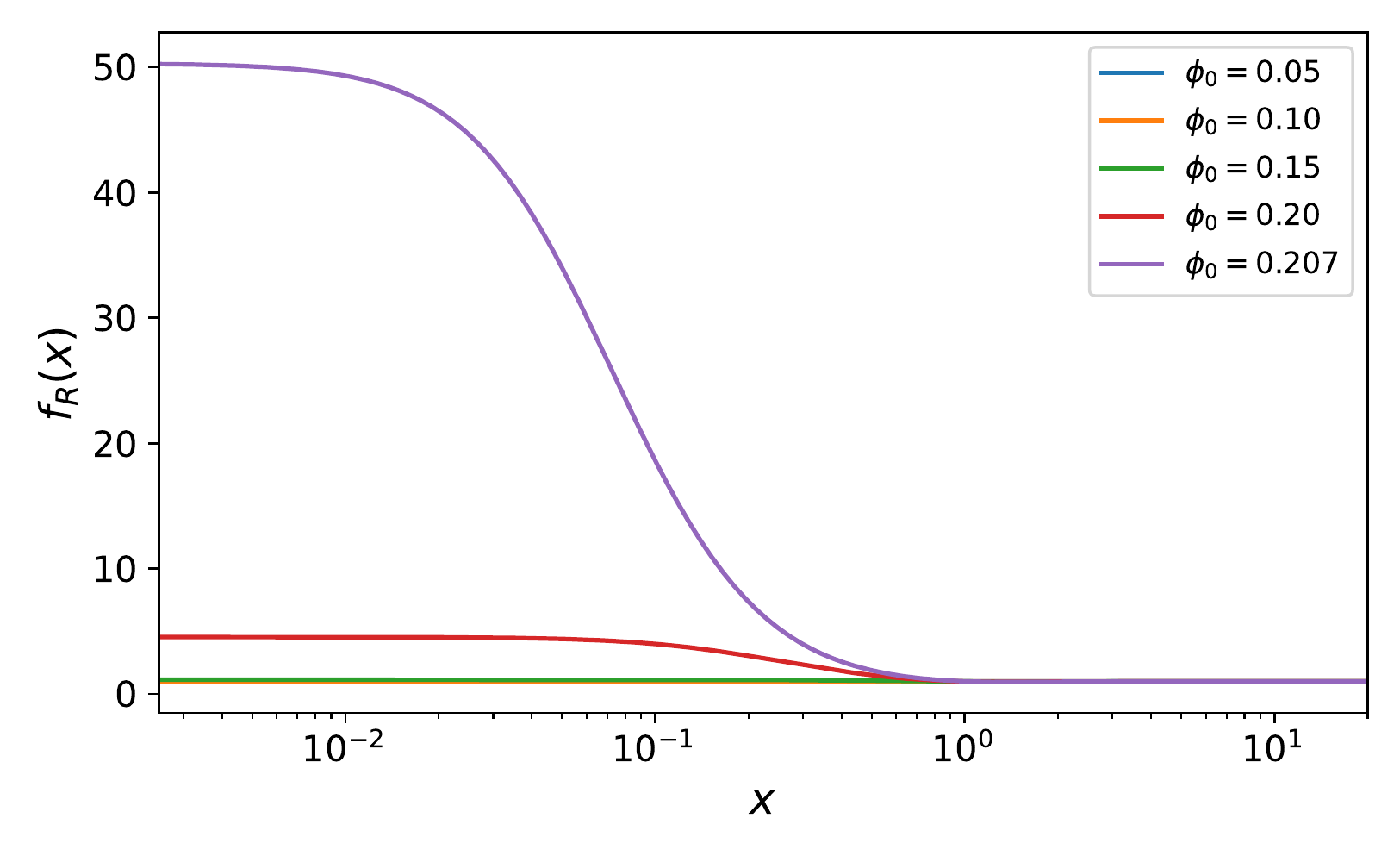}
	\caption{Radial profile of the conformal factor $f_R(x)$, for $\xi=-0.1$ for different boson star configurations. {Note that blue, orange, and green curves are overlapped.}}\label{fig:f_Rneg}
\end{figure}

\begin{figure}[h!]
	\centering
	\includegraphics[width=1.02\linewidth]{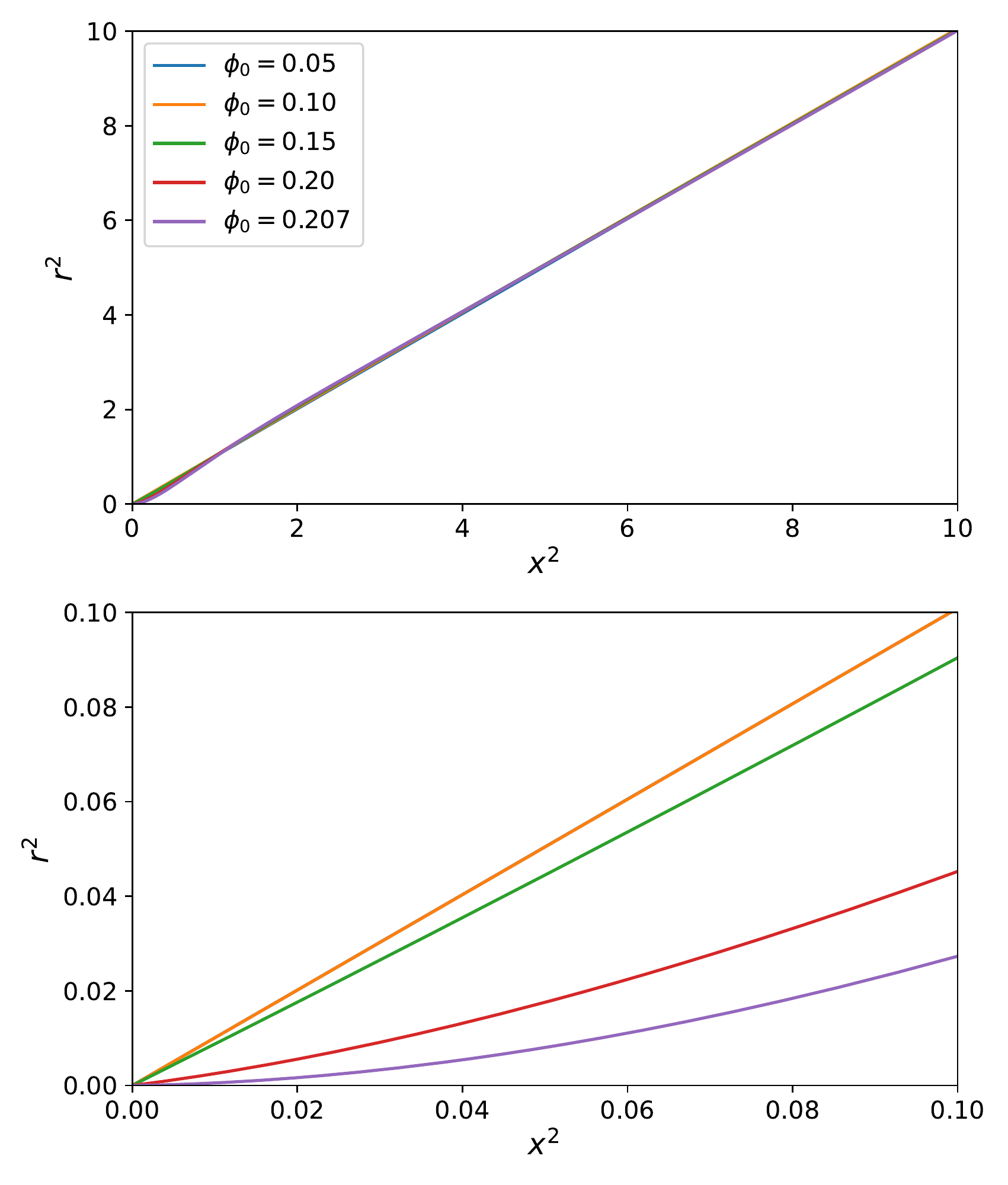}
	\caption{Relation between the area of the 2-spheres in both frames, for  $\xi=-0.1$. Notice the linearity between the areas in the two frames for most of the domain. The lower panel shows a zoom close to the origin where the linear relation is not satisfied. {Note that blue and orange curves overlap.}}
	\label{fig:radial-}
\end{figure}

In Appendix \ref{appendixA} we show radial plots of the metric components and of the scalar field for different values of $\phi_0$ in both frames to facilitate their comparison. From those plots one can see that all the relevant functions are smooth and show no divergences, not even for values close to the critical condition $\phi^2_0 \omega^2/\alpha^2_0 = -1/(2 \xi \kappa)$ {(see the purple curve in the plots)}. A close look at the figures in the $f(\mathcal{R})$ frame reveals that  the two metric functions $A^2(r)$ and $B^2(r)$ tend to zero at the origin  as the critical condition is approached. This is reasonable due to the appearance of the conformal factor  $f_\mathcal{R}$ in the denominator of those quantities, which has a rapid growth as one approaches the last point in the existence curve, $\phi_0\to \phi_0^{\text{last}}$, as shown in figure \ref{fig:f_Rneg}. Similarly, in figure \ref{fig:radial-} we see that the relation between the area of the spherical sectors in the two frames significantly deviates from linearity near the origin as one approaches this critical condition.  The observed flattening of $r^2(x)$ is similar to what happens in other Palatini models (coupled to electric fields) in which wormhole solutions arise \cite{Olmo:2015axa,Bambi:2015zch,Bejarano:2017fgz,Guerrero:2020uhn}. Unfortunately, our numerical exploration of the parameter space has not led to any satisfactory wormhole solution, with a minimum in $r(x)$ followed by a bounce, like those found in \cite{Afonso:2019fzv}, in which the mapping method was used in combination with the static, massless scalar field solution of GR as seed to generate new exotic compact objects in the same $f(\mathcal{R})$ theory as studied here. We suspect that the impossibility of finding that type of solutions in our analysis is due to the incompatibility of our boundary conditions at the center with those required to produce a bounce in the radial function $r(x)$. This interesting possibility will be further explored elsewhere. 

Let us now focus on the $\xi>0$ case in Fig.~\ref{fig:inidata}. In this case, as the gravitational coupling increases we see that features such as the local minimum or the spiraling occur earlier than in GR. This can be intuitively justified by the fact that the positive contribution of the quadratic curvature terms in the Lagrangian increase the gravitational attraction as compared to GR. For this reason, for a given central field amplitude $\phi_0$, the corresponding solution supports less mass than in GR, which justifies why $M_{\text{max}}$ is slightly lower than in GR. Correspondingly, for the same scalar field density, a higher deformation of spacetime is achieved as compared to GR. 

\begin{figure}
	\centering
	\includegraphics[width=\linewidth]{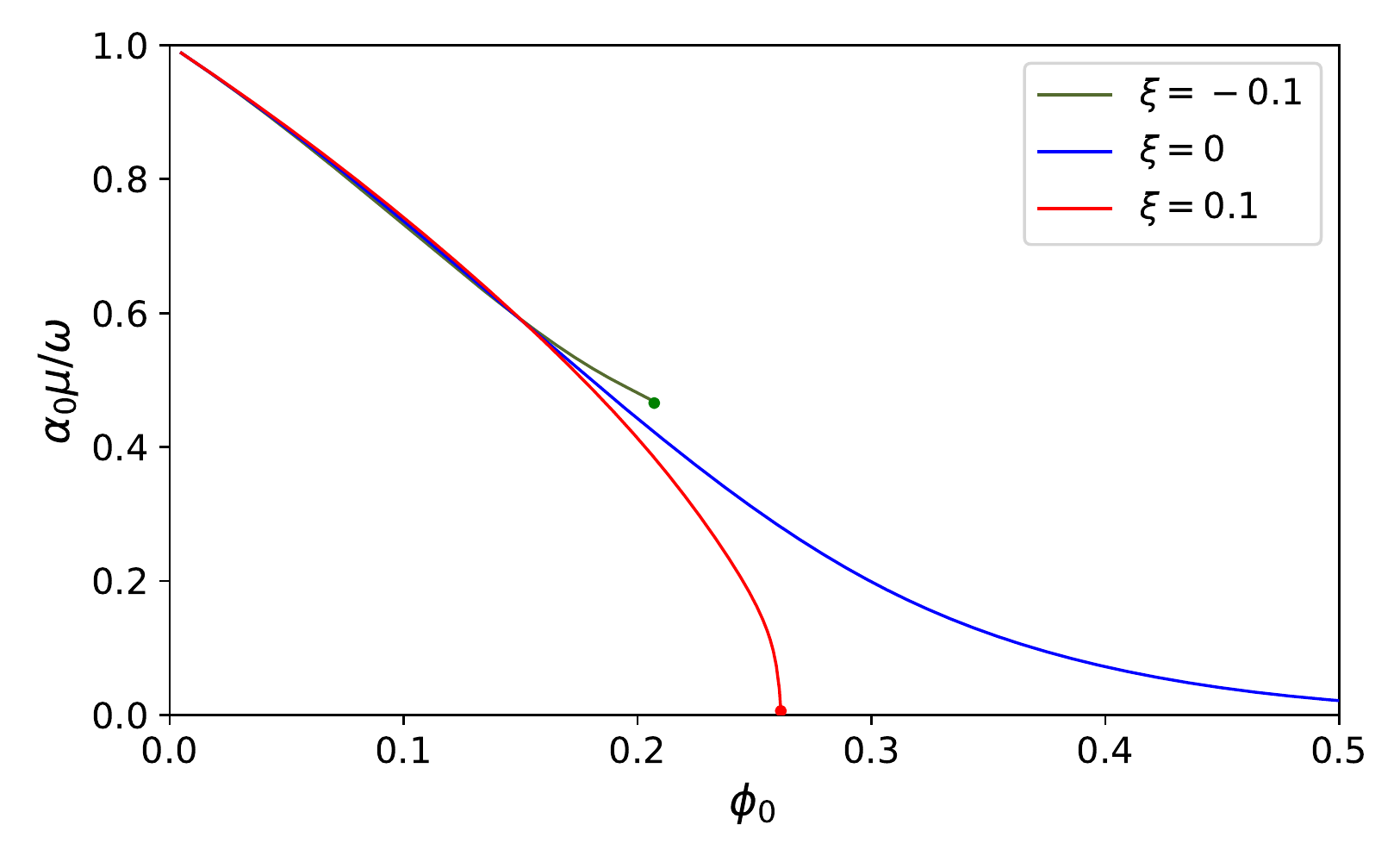}
	\caption{Metric parameter $\alpha_0$ evaluated at the origin vs $\phi_0$  for three different  values of $\xi$.}
	\label{fig:alpha_phi}
\end{figure}

\begin{figure}
	\centering
	\includegraphics[width=\linewidth]{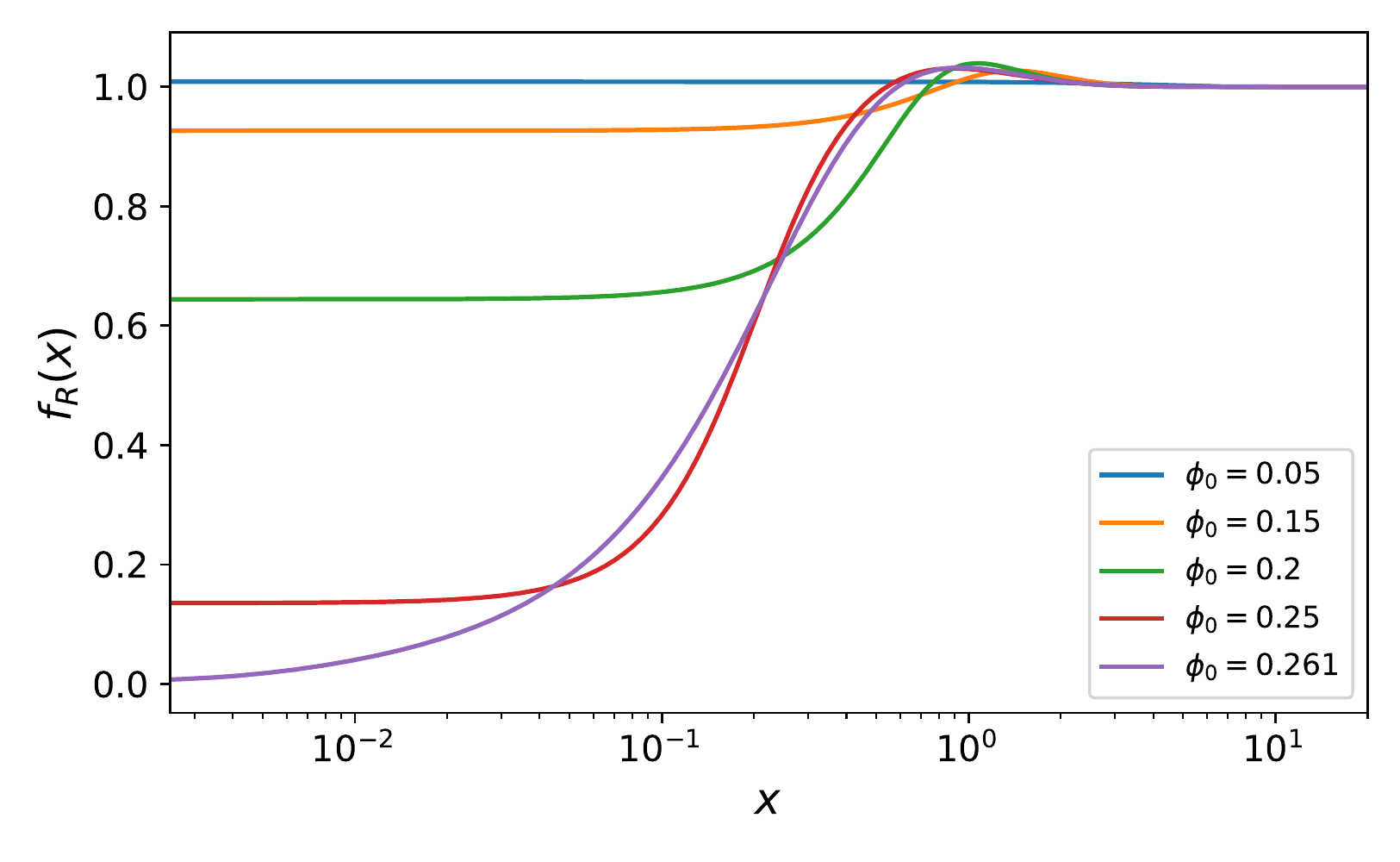}
	\caption{Radial profile of the conformal factor $f_R(x)$, for $\xi=0.1$ for different boson star configurations. }\label{fig:fRp}
\end{figure}

\begin{figure}
	\centering
	\includegraphics[width=1.02\linewidth]{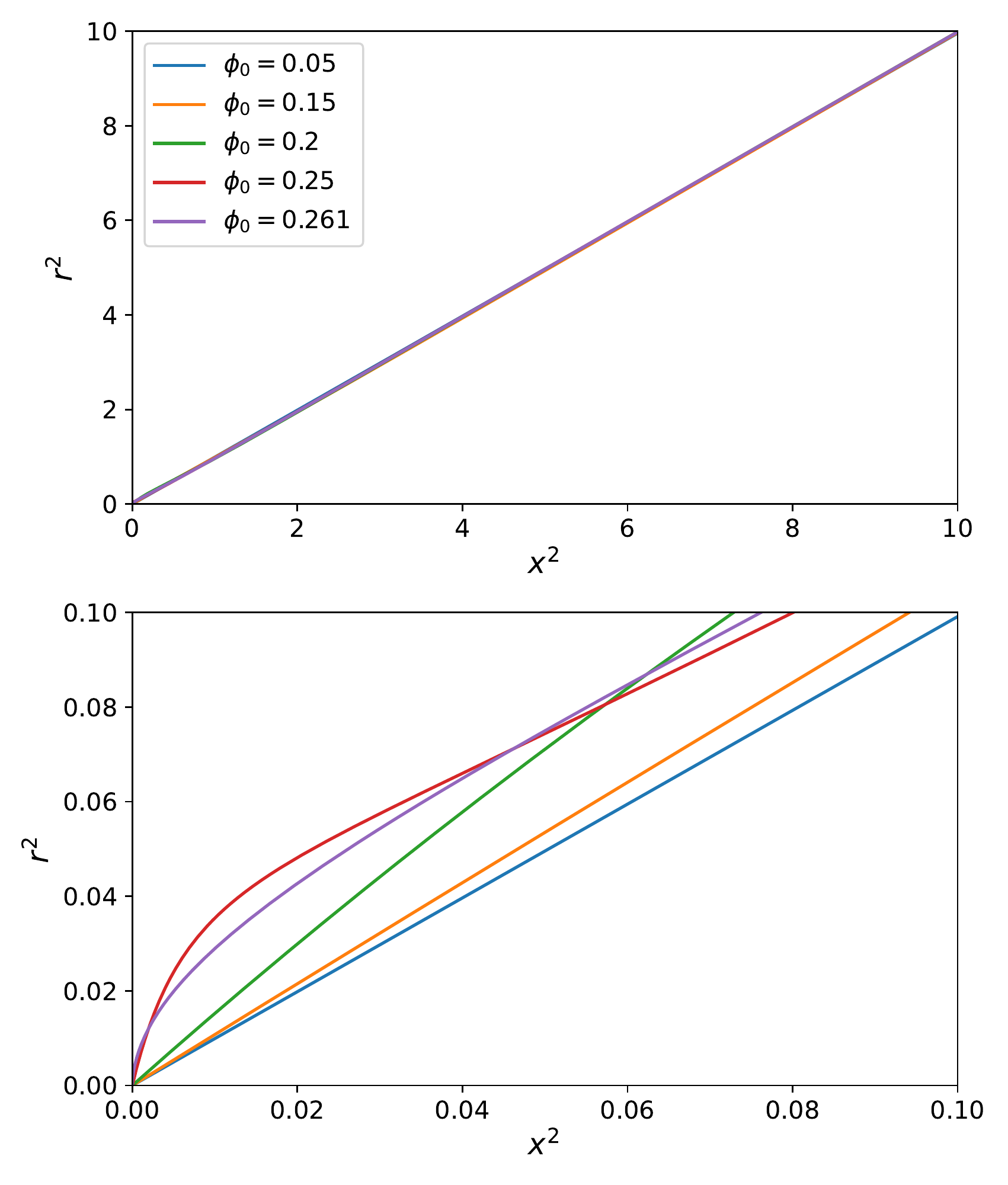}
	\caption{Relation between the area of the 2-spheres in both frames, for  $\xi=0.1$. As in Fig.~\ref{fig:radial-}
	the linearity between the areas in the two frames is broken only close to the origin, as shown in the lower panel. }
	\label{fig:radial+}
\end{figure}

The characteristic values of the last solutions we can build for the $\xi>0$ case are also reported on Table \ref{tab:values_inidata}. In this case, we are unable to find further solutions because equations \eqref{beta} and \eqref{psi} diverge at the origin. The reason for the divergence is that $\alpha_0$ tends to zero, as shown in Fig.~\ref{fig:alpha_phi}. 
We note that while in GR $\alpha_0$ tends to zero 
asymptotically, in  $f(\mathcal{R})$ with  $\xi>0$ it tends to zero abruptly. In analogy with the $\xi<0$ case, in Appendix \ref{appendixB} we provide plots of the radial profiles of the metric components and of the scalar field for $\xi>0$.  The corresponding figures show 
how the divergences at the origin in Eqs.~\eqref{beta} and \eqref{psi} as $\phi_0\to \phi_0^{\rm{last}}$ translate into a diverging $B^2(r)$ function. It is worth noting that even though $\alpha^2_0$ tends to zero for $\phi_0^{\rm{last}}$, its conformally related function $A^2_0$ tends to the finite value $0.204$. This  confirms that the conformal factor $f_\mathcal{R}$ also tends to zero at the origin when $\phi_0\to \phi_0^{\rm{last}}$ at the same rate as $\alpha_0$ (see Fig.~\ref{fig:fRp}). Finally, figure  \ref{fig:radial+} reveals that also in the $\xi>0$ case the linearity between the area of the 2-spheres in the two frames breaks near the center. However, contrary to the $\xi<0$ case (cf.~Fig.~\ref{fig:radial-}) it is now the curve $x=x(r)$ which flattens. A similar behavior has already been observed in other ECOs and might be related with the existence of wormhole structures in the Einstein frame geometry \cite{Afonso:2019fzv}, although in those cases $r^2(x)$ vanishes at a non-zero value of $x$.  

From the above discussion, we conclude that the amplitude of the gravitational coupling parameter $\xi$ magnifies the disparity of the boson star solutions in $f(\mathcal{R})$ with respect to GR, keeping the qualitative features of the solutions  essentially unchanged for low central densities but showing a trend towards new structures already seen in scenarios with wormholes at higher densities. A dedicated analysis of this trend at high central field amplitudes will be carried out elsewhere, where non-standard boundary conditions will be implemented.

\subsection{Self-interacting scalar field}

The results obtained in the previous section for a free massive scalar field can be extended to the case in which self-interactions are present, as we discuss next. By considering a potential of the form
\begin{equation}
	V_{\text{SI}}=-\frac{1}{2}\mu^2 |\Phi|^2-\frac{1}{4}\lambda |\Phi|^4\quad ,
\end{equation}
in which the second term is a quartic self-interaction term, 
the system of differential equations to build the stellar models is modified as follows:
\begin{equation}\label{beta_si}
\begin{aligned}
\frac{\partial_{x}\beta}{\beta}&=\frac{1-\beta^2}{2x}\\
&+\frac{1}{1+4 \xi \kappa \left( \mu^2 \phi^2 +\frac{\lambda \phi^4}{2} \right)}\frac{\kappa x}{4} \left\lbrace 
\beta^2   \left( \mu^2 \phi^2 +\frac{\lambda \phi^4}{2} \right) \right.\\
&\left.+\left( \omega^2\phi^2\frac{ \beta^2  }{ \alpha^2}+ \psi^2 \right)\left[1-2 \kappa \xi\left( -\frac{\omega^2   \phi^2}{ \alpha^2} +\frac{\psi^2}{ \beta^2} \right)\right] \right.\\
& \left.+2\kappa \xi \beta^2\left( \frac{\omega^2   \phi^2}{ \alpha^2} -\frac{\psi^2}{ \beta^2} \right)^2  \right\rbrace ,  
\end{aligned}
\end{equation}
\begin{equation}\label{alpha_si}
\begin{aligned}
\frac{\partial_{x}\alpha}{\alpha}&=\frac{\beta^2-1}{x}+\frac{\partial_{x}\beta}{\beta}\\
&+\frac{1}{1+4 \xi \kappa  \left( \mu^2 \phi^2 +\frac{\lambda \phi^4}{2} \right) }\frac{\kappa x}{4}\left\lbrace\frac{}{}-2  \beta^2  \left( \mu^2 \phi^2 +\frac{\lambda \phi^4}{2} \right) \right.\\
&\left.-2\kappa \xi \beta^2\left( \frac{\omega^2   \phi^2}{ \alpha^2} -\frac{\psi^2}{ \beta^2} \right)^2\right\rbrace .
\end{aligned}
\end{equation}
\begin{equation}\label{psi_si}
\begin{aligned}
\partial_{x}\psi&=\frac{1}{\left[1+4 \xi \kappa  \left( \mu^2 \phi^2 +\frac{\lambda \phi^4}{2} \right) \right]\left[1-2 \xi \kappa\left( -\frac{\omega^2 \phi^2 }{\alpha^2}+  \frac{3\psi^2}{\beta^2}\right)\right]} \left\lbrace\frac{}{} \right.\\ 
&-\psi\left(\frac{2}{x}+\frac{\partial_{x}\alpha}{\alpha}- \frac{\partial_{x}\beta}{\beta} \right)\left[1+4\kappa \xi  \left( \mu^2 \phi^2 +\frac{\lambda \phi^4}{2} \right)\right]\\
&\times\left[1-2 \xi \kappa\left( -\frac{\omega^2 \phi^2 }{\alpha^2}+  \frac{\psi^2}{\beta^2}\right)\right]\\
&-\omega^2 \phi \frac{\beta^2}{\alpha^2}\left[1+2 \xi \kappa\left( \frac{\omega^2 \phi^2 }{\alpha^2}+  \frac{\psi^2}{\beta^2}\right)\right]\\
&+\beta^2 \left( \mu^2 \phi +\lambda \phi^3 \right) \left(1+4\kappa \xi \frac{\psi^2}{\beta^2}\right)\\
&+\kappa \xi \left[ \frac{4 \omega^2 \phi^2 \psi}{\alpha^2}\frac{\partial_{x}\alpha}{\alpha}\left(1+4\kappa\xi \left( \mu^2 \phi^2 +\frac{\lambda \phi^4}{2} \right)\right)\right.\\
&-\frac{4\psi^3}{\beta^2}\frac{\partial_{x}\beta}{\beta}\left(1+4\kappa \xi  \left( \mu^2 \phi^2 +\frac{\lambda \phi^4}{2} \right)\right)\\
&\left.+2  \phi^5 \omega^2 \lambda  \frac{\beta^2}{\alpha^2}\right]\\
&+\kappa^2 \xi^2\left[-12 \left(\mu^2+\lambda \phi^2\right) \phi \frac{ \psi^4}{\beta^2}\right.\\
&\left.\left.-4 \mu^2 \frac{\beta^2\omega^4\phi^5}{\alpha^4}+4 \lambda   \frac{ \omega^2\psi^2 \phi^5 }{\alpha^2}\right]\right\rbrace.\\
\end{aligned}
\end{equation}
For the numerical analysis it is convenient to re-scale the self-interaction parameter as  
\begin{equation}
	\Lambda=\frac{2}{\kappa}\lambda \quad. 
\end{equation}

\begin{figure}[t]
	\includegraphics[width=\linewidth]{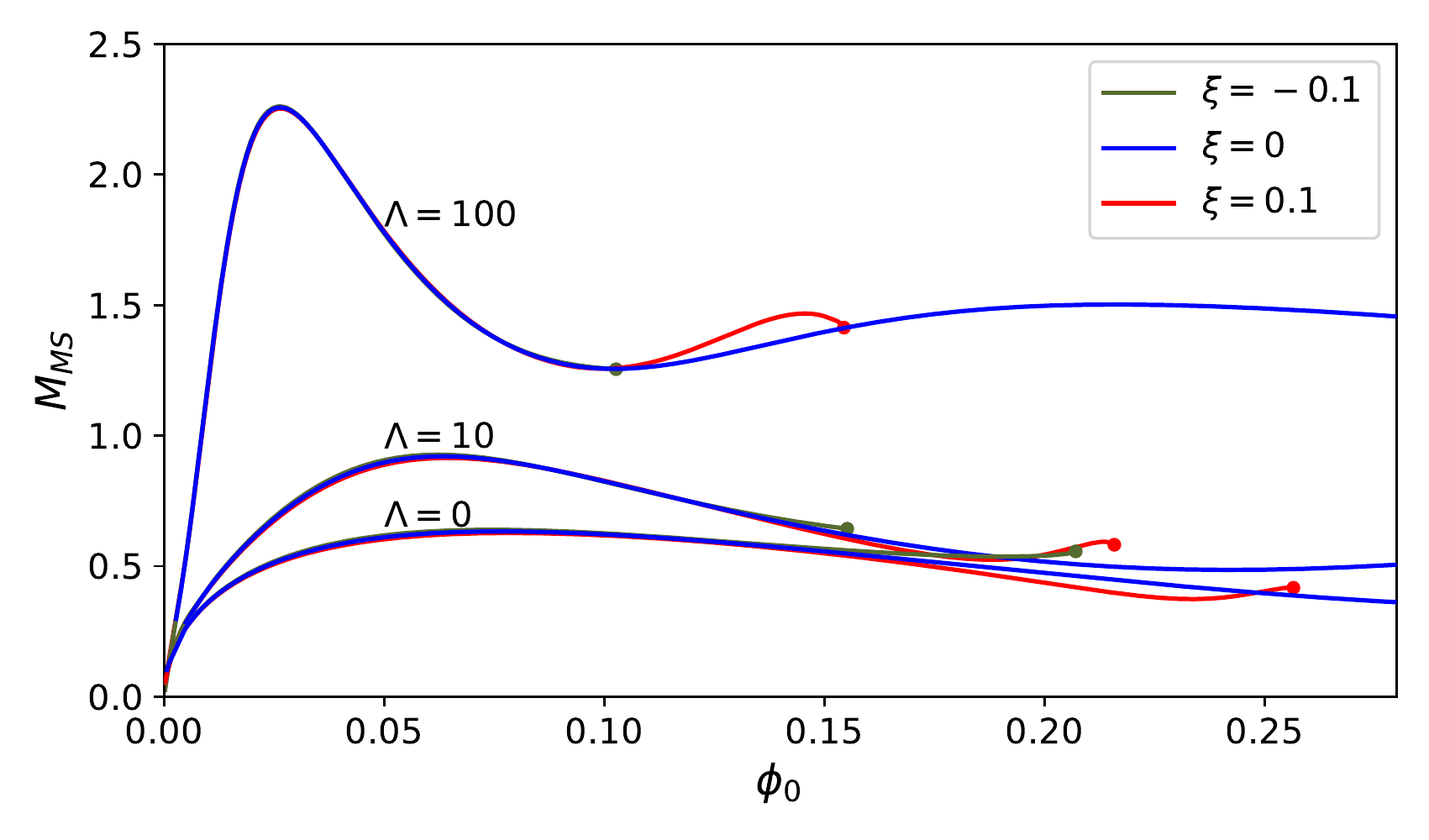}
	\includegraphics[width=1.03\linewidth]{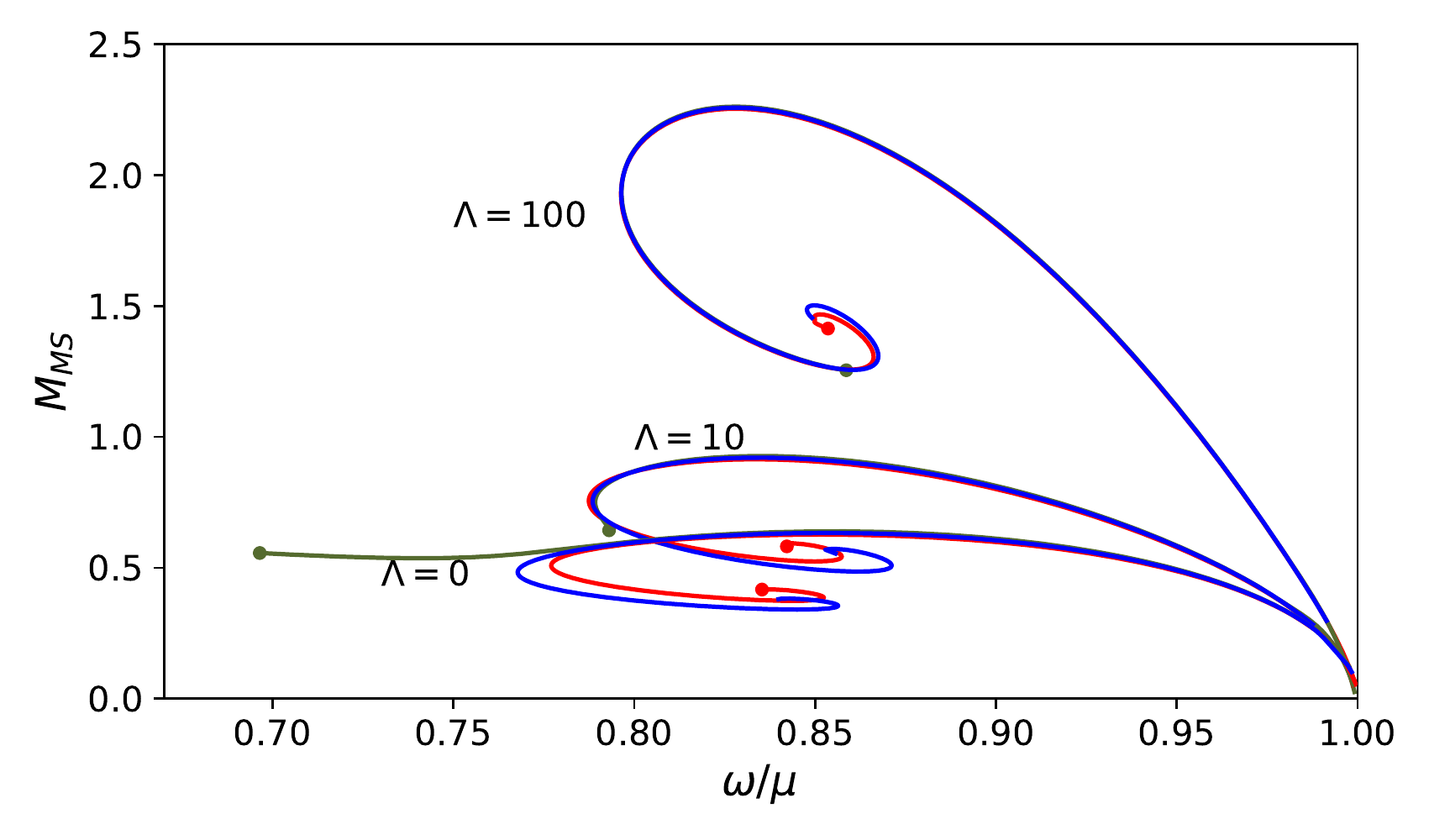}
	\caption{Equilibrium configurations of boson stars in Palatini $f(\mathcal{R})$ gravity for three different values of the coupling parameter $\xi$ and of the self-interaction parameter $\Lambda$.
Top panel: total mass as a function of the central value of the scalar field. Bottom panel: total mass as a function of the frequency.}
	\label{fig:selfin}
\end{figure}

The corresponding boson star solutions are plotted in figure \ref{fig:selfin}. This figure shows existence curves for different values of the self-interaction parameter $\Lambda$ in GR and in Palatini $f(\mathcal{R})$ gravity.  We only consider  $\Lambda\ge0$ because a negative value would violate energy conditions.
Figure \ref{fig:selfin} exhibits that 
increasing the self-interaction results in more massive boson stars and in a larger maximum mass (see also~\cite{escorihuela2017quasistationary}). Moreover, this figure reveals that the  existence curves shorten as the self-interaction parameter $\Lambda$ increases. For $\xi<0$, $\alpha_0$ decreases faster as $\Lambda$ increases producing the observed shortening (compare the location of the green circles in the figure with those of the red and blue circles). Paying attention to the green curves in figure \ref{fig:selfin}, corresponding to $\xi=-0.1$, one can see that for $\Lambda=10$ the disparity with GR is hardly noticeable and for $\Lambda=100$ the existence curve lays over the GR curve making its length the only remarkable difference.
Meanwhile for $\xi=0.1$ (red curves) the curves are shorter by the fact that the conformal factor now goes as
\begin{equation}
f_{\mathcal{R}}=\frac{1+4\xi \kappa \mu^2 \phi^2+2 \xi \kappa \lambda \phi^4}{1-2\xi \kappa \left(-\frac{\omega^2 \phi^2}{\alpha^2}+\frac{\psi^2}{\beta^2}\right)},
\end{equation}
making the condition $f_\mathcal{R}=0$ easier to achieve.

\bigskip

\section{Einstein frame perspective}
\begin{figure}[t]
	\includegraphics[width=\linewidth]{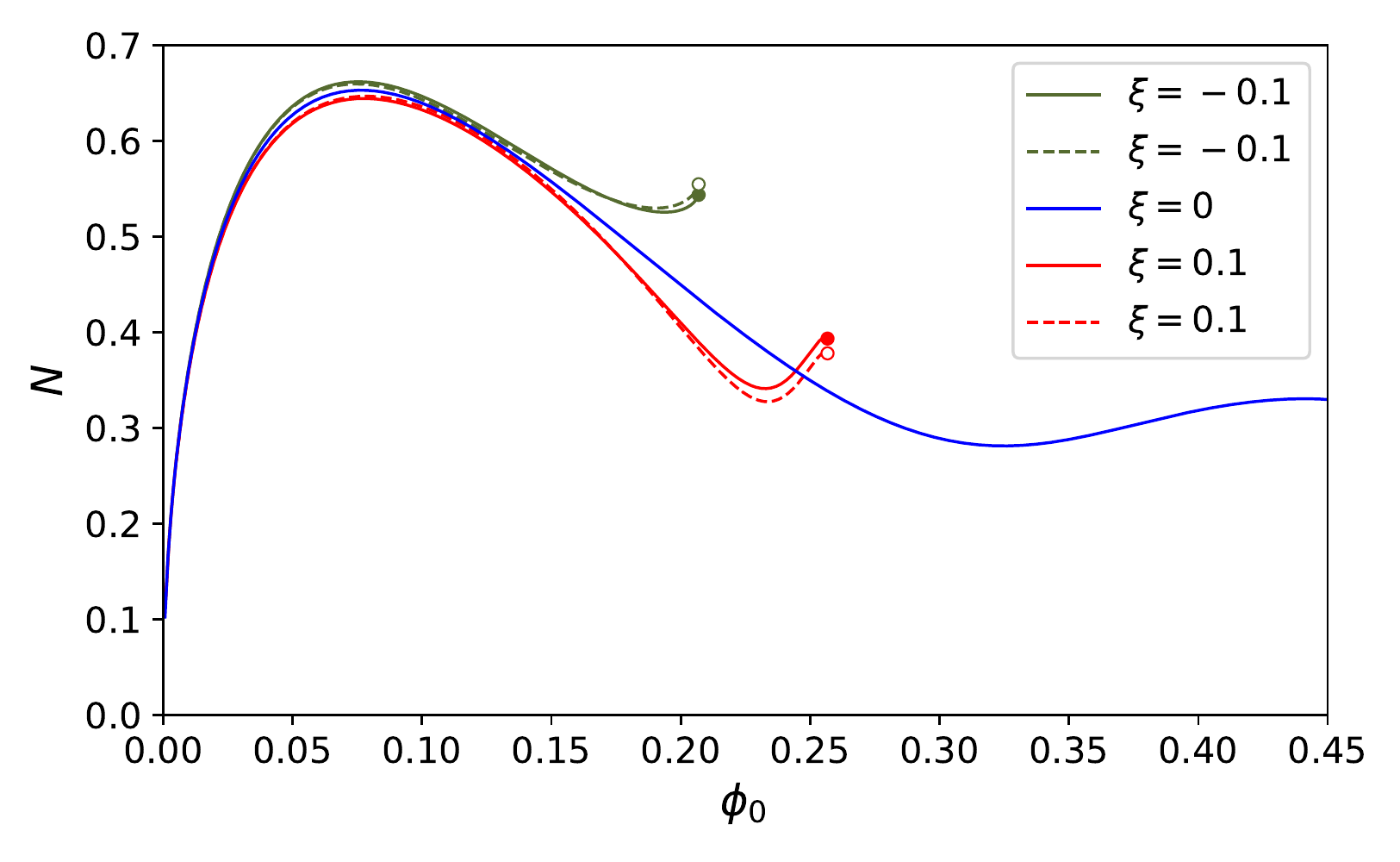}
	\includegraphics[width=\linewidth]{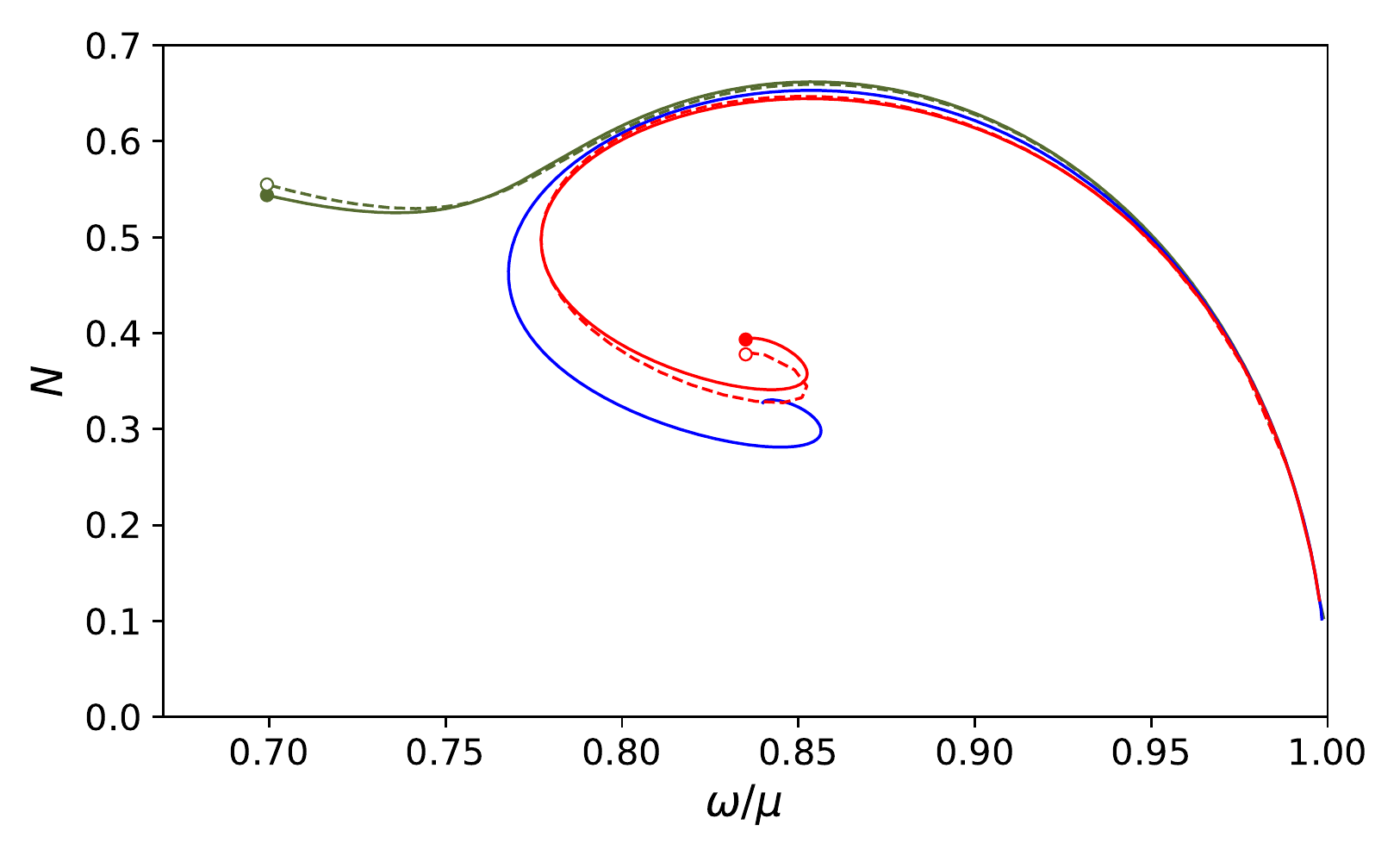}
	\caption{Equilibrium configurations of {canonical boson star matter lagrangian} in Palatini $f(\mathcal{R})$ gravity (solid lines) and {modified matter lagrangian} in GR (dashed lines). Top panel: particle number as a function of the central value of the scalar field. Bottom panel: same quantity but plotted against the frequency. Dashed curves represent $N_{EF}$ as defined in Eq. (\ref{eq:N_EF}) while solid curves correspond to $N$ as defined in Eq. (\ref{eq:N_fR}).}
	\label{fig:N_EF}

\end{figure}

{The computational approach followed in this work to analyze boson stars in Palatini $f(\mathcal{R})$ gravity requires first to build a boson star solution in GR generated by an exotic matter source (see Eq.~(\ref{eq:K})), which consists on a $K$-essence piece \cite{ArmendarizPicon:2000ah} plus a modified potential term. As already pointed out below Eq.~(\ref{eq:M_MS}), the mass corresponding to the boson star in the Einstein frame theory is essentially the same as that in the $f(\mathcal{R})$ frame. The scalar field amplitude and frequency are independent of the frame, which implies that the existence curves of Fig.~\ref{fig:inidata} representing $M_{\rm MS}$ as a function of $\phi_0$ and $\omega$ are also valid for the boson star of the Einstein frame theory of Eq.~(\ref{eq:K}). The number of particles, however, is expected to be different. The conserved charge corresponding to the non-linear Lagrangian (\ref{eq:K}) takes the form}
\begin{eqnarray}\label{eq:N_EF}
N_{\rm{EF}}&=&\int_{\Sigma}dV\sqrt{-q}q^{t \nu} \frac{i}{2}K_Z\left(\bar{\Phi}\partial_{\nu}\Phi-\Phi\partial_{\nu}\bar{\Phi}\right) \nonumber \\
&=&4\pi\int_{0}^{\infty} dxx^2 \frac{ \omega\phi^2 \beta}{\alpha} K_Z \ ,
\end{eqnarray}
{where $K_Z=\partial K/\partial Z$. With elementary algebra, one can see that for our $f(\mathcal{R})$ theory $K_Z$ coincides with $1/f_\mathcal{R}$, which eventually turns Eq.~(\ref{eq:N_EF}) into}
\begin{eqnarray}
   N_{\rm EF}= 4\pi\int_{0}^{\infty} dxx^2 \frac{ \omega\phi^2 \beta}{ \alpha f_\mathcal{R}}  \ .
\end{eqnarray}
{We thus see that this expression simply differs from that for the particle number in the $f(\mathcal{R})$ frame, Eq.~(\ref{eq:N_fR}), in the power of $f_\mathcal{R}$ in the denominator. The numerical result for this quantity is displayed in Fig.~\ref{fig:N_EF}. This figure shows that for a given value of $\xi$, the existence curves for GR with modified matter (dashed lines) are strongly degenerate with $f(\mathcal{R})$ coupled to canonical matter (solid lines) over all the domain of existence, following almost identical paths in the space of solutions. We also see that for small values of the central scalar field amplitude or for large frequencies, the results are almost coincident with those of GR with canonical matter, with noticeable differences arising only as the last solution of each branch is approached. 
Interestingly, the range of values in which solutions can be found is limited also in the GR case with modified potential. The reason for this is apparent from the fact that algebraically $K_Z$ coincides with $f_\mathcal{R}$. Thus, whenever the conformal factor $f_\mathcal{R}$ has a problem (either vanishes or diverges), $K_Z$ will also have problems. This is the main reason why the range of existence of solutions in the model studied here is shorter than in other modified theories of gravity \cite{Baibhav:2016fot,Brihaye:2016lin,Ilijic:2020vzu}.} 

\section{Conclusion}
\label{conclusions}

{In this work we have built numerical solutions for spherically symmetric, static boson stars in the quadratic Palatini theory $f(\mathcal{R})=\mathcal{R}+\xi \mathcal{R}^2$. The complex scalar field that generates the solutions is characterized by a canonical Lagrangian with a mass, a non-zero frequency, and a self-interaction term of the $|\Phi|^4$ type. Our numerical approach took advantage of the correspondence \cite{Afonso:2019fzv} that exists between this type of gravity theories and GR, such that the original modified gravity theory can be turned into a modified matter theory coupled to standard GR. We have thus solved the corresponding Einstein-Klein-Gordon system of equations and used the results to construct the solutions of the $f(\mathcal{R})$ theory. }

{The main results depicting the corresponding existence curves of the $f(\mathcal{R})$ theory are shown in Fig.~\ref{fig:inidata}. An important difference with respect to GR is the limited range of scalar field amplitudes allowed at the center of the star, which is much shorter than in GR (see also  Fig.~\ref{fig:alpha_phi}). For relatively small central field amplitudes $\phi_0$, we found that the solutions do not differ significantly from those of GR, though larger/smaller masses can be accommodated for a given $\phi_0$ depending on whether $\xi<0$ or $\xi>0$, respectively. New features arise in the $\xi<0$ case regarding the dependence of the total mass and particle number of a solution with its oscillation frequency. In GR coupled to canonical matter, these curves exhibit a characteristic spiral pattern which is lost in this case (at least) in the range of parameters explored (see the upper right panel of Fig.~\ref{fig:inidata}). Although other theories of gravity may also depart from this spiral pattern \cite{Baibhav:2016fot,Brihaye:2016lin,Ilijic:2020vzu}, the model considered here is peculiar because the range over which solutions are possible is relatively small, which could facilitate its observational discrimination.}

{Though our focus was on analyzing canonical boson stars coupled to $f(\mathcal{R})$ gravity, the fact is that our computational method forced us to construct boson star solutions in GR coupled to unconventional matter [see Eq.(\ref{eq:K})]. In this regard, we note that the total mass, field amplitudes, and frequencies that we obtained are valid in both theories, namely, in $f(\mathcal{R})$ coupled to the scalar Lagrangian $P(X,\Phi)=X-2V(\Phi)$ and in GR coupled to the non-canonical scalar Lagrangian $K(Z,\Phi)$. It is also easy to see from Figs.~\ref{fig:radial-} and \ref{fig:radial+} that the radius of these stars will also be practically indistinguishable because the conformal factor that relates the radial coordinates $r$ and $x$ is essentially equal to unity at the surface. More explicitly, Fig.~\ref{fig:N_EF} shows that the number of particles for GR and $f(\mathcal{R})$ not only follow the same trend and have the same domain of definition, but also that they are almost coincident over the whole range of solutions. All this puts forward an interesting degeneracy between boson stars in GR coupled to exotic matter and in $f(\mathcal{R})$ coupled to standard (or canonical) matter. We note also, in this sense, that the incorporation of self-interactions in the scalar field potential (see Fig.~\ref{fig:selfin}) simply contributes to worsen this degeneracy. It is thus necessary to go beyond the basic setup considered here in order to determine if other observables could help break this degeneracy. Among other possibilities, the consideration of scenarios with less symmetry, such as axially symmetric, rotating solutions, the stability under small perturbations, or the coupling to non-scalar matter fields may offer relevant information to distinguish between GR and $f(\mathcal{R})$ theories.  The exploration of more general Palatini theories is also important to better understand the extent and underlying reason of the observed  degeneracies. Note, in this sense, that $f(\mathcal{R})$ theories are simply conformally related to GR, whereas models such as $R+\xi R^2+\lambda R_{\mu\nu}R^{\mu\nu}$ or of Born-Infeld type involve disformal transformations between the original and the corresponding Einstein frames \cite{Afonso:2018hyj,Delhom:2019zrb,Afonso:2018mxn,Afonso:2018bpv,Orazi:2020mhb}. This implies more deformation functions than just a conformal factor, which could have an impact in the range of definition and shape of the existence curves.  }

{As a final remark, we note that as one approaches the limiting value of the central field amplitude, the conformal factor strongly deviates from unity (either towards infinity if $\xi<0$ or towards zero if $\xi>0$) and leads to a deformation of the relation between areas of the $f(\mathcal{R})$ and GR frames (see Figs.~\ref{fig:radial-} and \ref{fig:radial+}) that reminds of the $r(x)$ relation that appears in scenarios with wormholes such as those described in \cite{Afonso:2019fzv}. We suspect that different boundary conditions than those considered here for boson stars might be necessary in order to accommodate a throat (minimum in $r(x)$). We hope to report on this and other issues in future works. }

\acknowledgements
 AMF is supported by the Spanish Ministerio de Ciencia y Innovaci\'on with the PhD fellowship PRE2018-083802. NSG is supported by the Center for Research and Development in Mathematics and Applications (CIDMA) through the Portuguese Foundation for Science and Technology (FCT - Funda\c c\~ao para a Ci\^encia e a Tecnologia), references UIDB/04106/2020 and UIDP/04106/2020, and by the projects PTDC/FIS-OUT/28407/2017,  CERN/FIS-PAR/0027/2019 and PTDC/FIS-AST/3041/2020. JAF acknowledges support from the Spanish Agencia Estatal de Investigaci\'on (PGC2018-095984-B-I00) and by the Generalitat Valenciana (PROMETEO/2019/071).  GJO acknowledges support from the Spanish Agencia Estatal de Investigaci\'on (FIS2017-84440-C2-1-P), by the Generalitat Valenciana (PROMETEO/2020/079), and by the Spanish Research Council (i-COOPB20462, CSIC).
 This work has further been supported by the European Union's Horizon 2020 Research and Innovation (RISE) programme H2020-MSCA-RISE-2017  Grant  No.   FunFiCO-777740.

\clearpage
\newpage
\onecolumngrid
\bigskip
\appendix

\section{EKG system solutions with $\xi<0$}
\label{appendixA}

This appendix shows radial plots of the metric components and of the scalar field in both frames and for the case $\xi=-0.1$, to facilitate the description of the results shown in Section~\ref{results}. Results  for other negative values of $\xi$ are similar and, thus, they are not shown. The figures display models for different values of $\phi_0$. All functions show  smooth profiles and no divergences, not even for values of $\phi_0$  close to the critical condition $\phi^2_0 \omega^2/\alpha^2_0 = -1/(2 \xi \kappa)$ {(see purple curves)}.

\begin{figure}[h!]

    \includegraphics[width=0.75\linewidth]{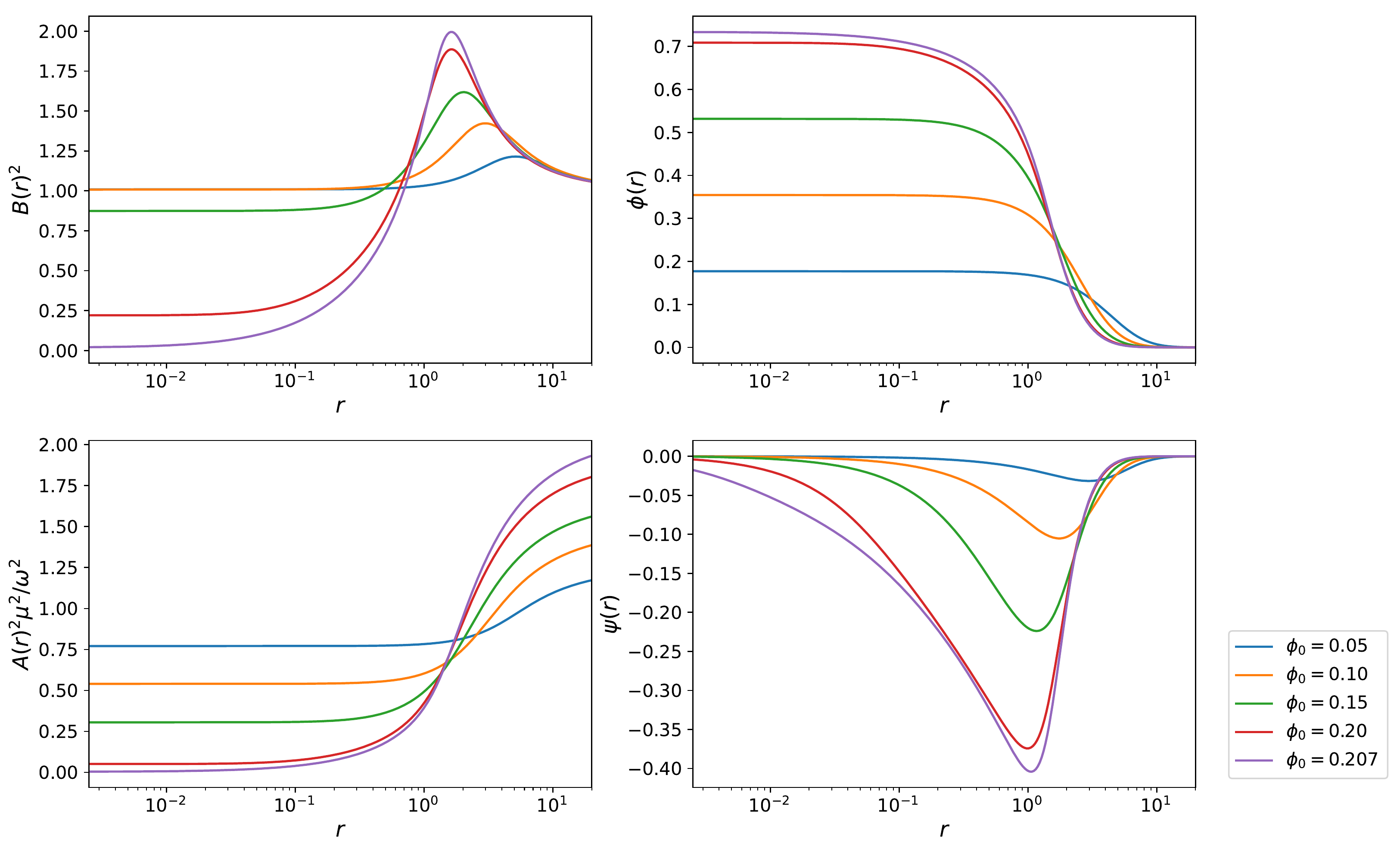}

    \vskip 1cm
    \includegraphics[width=0.75\linewidth]{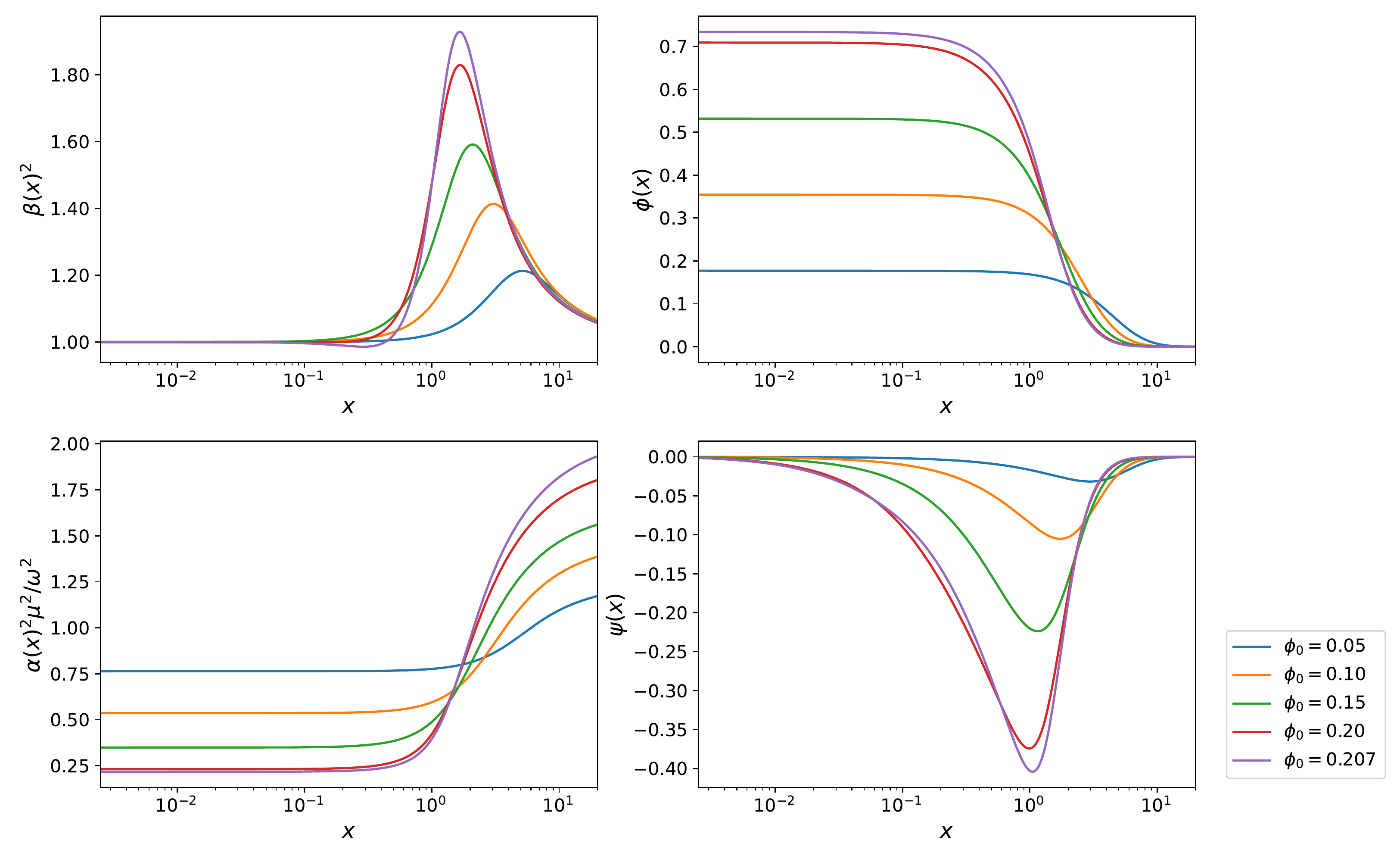}
	\caption{Radial profiles of the metric functions and of the scalar field functions in both frames for $\xi=-0.1$. Five models of boson stars are plotted, as indicated by the value of $\phi_0$ shown in the legend.} 
	\label{fig:fmet_dens_neg}
\end{figure}

\section{EKG system solutions with $\xi>0$}
\label{appendixB}
This appendix shows radial plots of the metric components and of the scalar field in both frames and for the case $\xi=0.1$, to facilitate the description of the results shown in Section~\ref{results}.

\begin{figure*}[h!]
	\centering
    \includegraphics[width=0.75\linewidth]{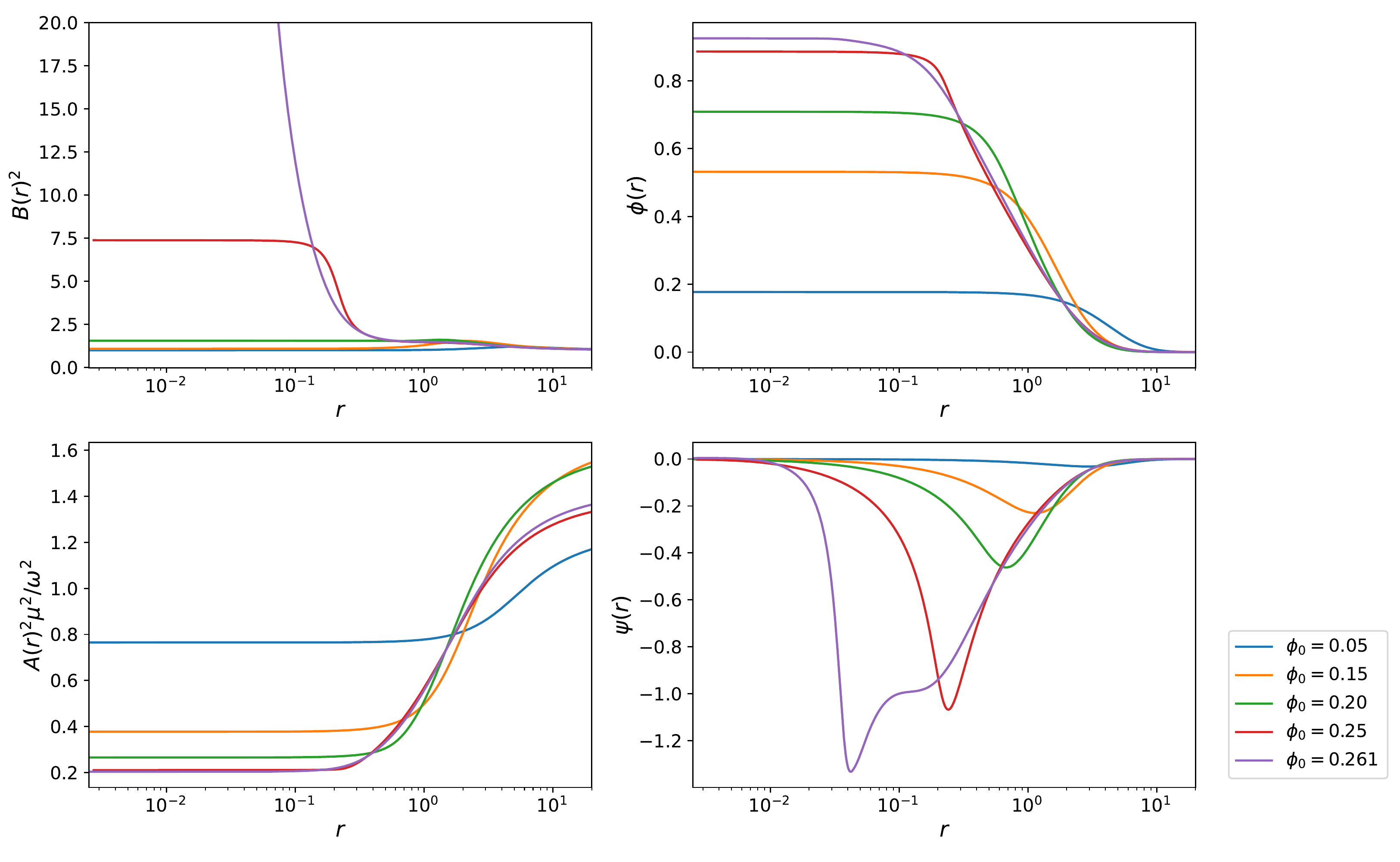}
    \vskip 1cm
    \includegraphics[width=0.75\linewidth]{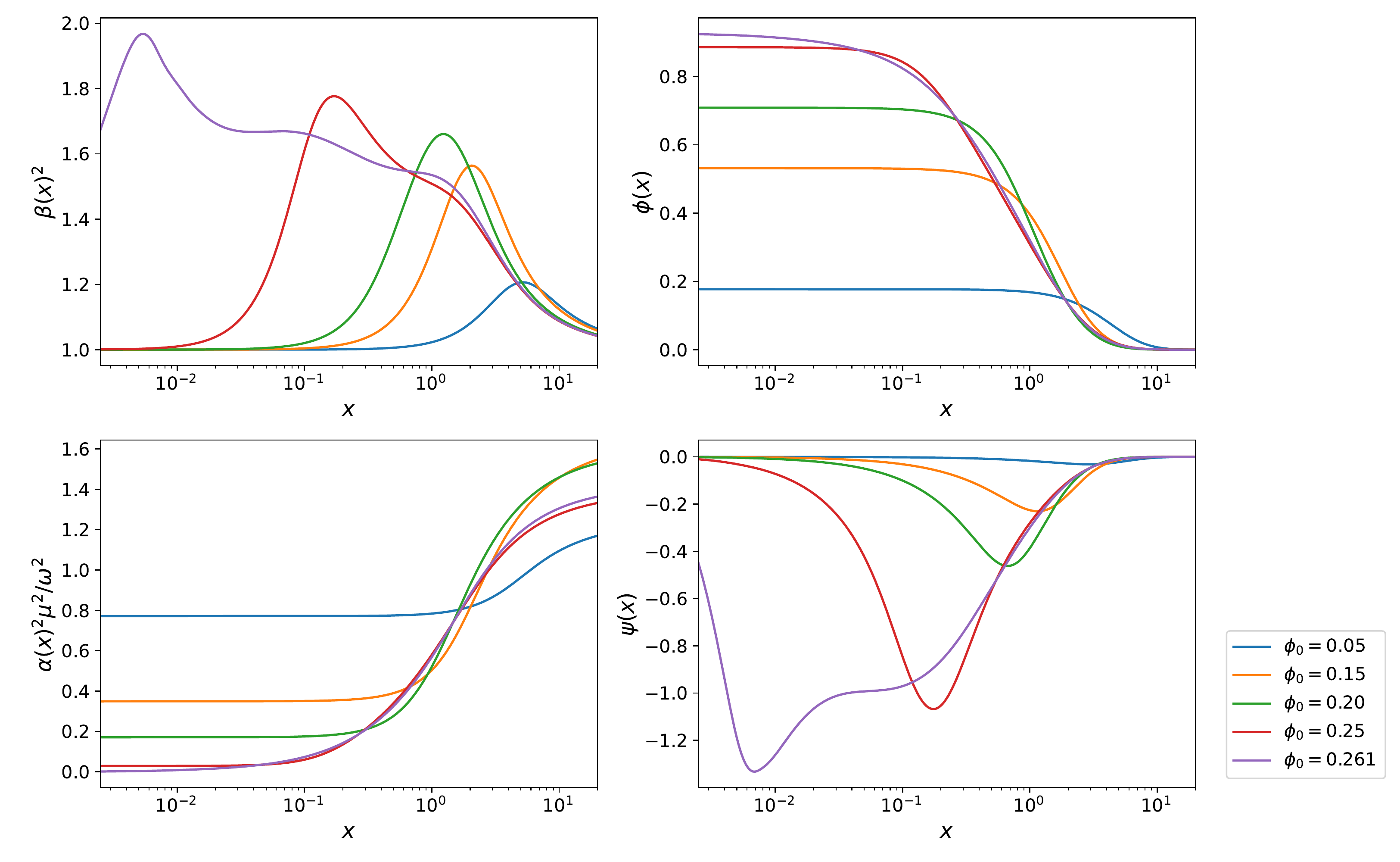}
	\caption{Radial profiles of the metric functions and of the scalar field functions in both frames for $\xi=-0.1$. Five models of boson stars are plotted, as indicated by the value of $\phi_0$ shown in the legend.} 
	\label{fig:fmet_dens_neg}
\end{figure*}

\end{document}